\documentclass[prl,twocolumn,aps,amsmath,amssymb]{revtex4}

\usepackage{graphicx}
\usepackage{dcolumn}
\usepackage{bm}
\usepackage{wrapfig}
\usepackage{amsmath,amsfonts,amssymb,bm,ulem}
\usepackage[usenames]{color}
\usepackage{bm}
\usepackage{slashed}
\usepackage{mathrsfs}
\usepackage{graphicx}
\usepackage{dcolumn}
\usepackage{bm}

\newcommand{\be}{\begin{equation}}
\newcommand{\ee}{\end{equation}}

\begin{document}

\title{A Functional approach to soft graviton scattering and BMS charges}

\author{Jordan Wilson-Gerow}
 \email{wilsonjs@phas.ubc.ca}
\author{Colby DeLisle}
 \email{cdelisle@phas.ubc.ca}
\author{Philip Stamp}
 \email{stamp@phas.ubc.ca}
\affiliation{
 Department of Physics \& Astronomy\\
 and\\
 Pacific Institute for Theoretical Physics\\
 University of British Columbia\\
 Vancouver, BC, V6T 1Z1, Canada
}

\begin{abstract}
We consider the interaction between a matter system and soft gravitons. We use a functional eikonal expansion to deal with the infrared divergences, and introduce a ``composite generating functional" which allows us to calculate a decoherence functional for the time evolution of the system. These techniques allow us to formulate scattering problems in a way which deals consistently with infrared effects, as well as being manifestly diffeomorphism invariant. We show how the asymptotic form of the decoherence functional can be written in terms of the infinitely many conserved charges associated with asymptotic BMS symmetries, and allow us to address the question of how much information is lost during the scattering.

\end{abstract}

\maketitle


\section{I: Introduction}
\label{sec:intro}


Although the black hole information paradox \cite{hawking76} has been with us now for over 4 decades, it is without any generally accepted resolution - recent reviews \cite{unruhW17,marolf17} indicate the depth of the issues involved. One idea that has emerged recently in this connection focusses on soft gravitons and soft photons, and the asymptotic charges associated with these \cite{stromSG,hawking16,hawking17,HPY14,ACS14,strom14,laddha15}. Insofar as black holes are concerned, the idea here is that information loss from the black hole will arise from both photon and graviton emission, and that this information is stored at the future boundary of the horizon. The information can, in this scenario, be described  in terms of ``charges" at future infinity; in the case of gravitons, these ``BMS charges" are associated with BMS supertranslations. Extensive discussions of this point of view appear in the recent papers of Strominger et al. \cite{hawking17,stromN,stromP}.

Quite apart from any implications for the physics of black holes, this work has raised important questions about the information loss associated with soft bosons coupled to matter fields: currently there is strong disagreement over whether there is any information loss at all, and if so how much.

One point of view argues that the emission of soft bosons, with its associated infrared catastrophe, must be associated with information loss - the information is carried off in the form of bremsstrahlung radiation, by an infinite number of soft bosons. According to this point of view, we must average over the soft bosons, noting that any information contained in them is only meaningful if one can access it using some measuring system, which will inevitably have a finite energy discrimination (typically formulated in terms of an IR cutoff on the boson excitations). This point of view goes back to early formulations of the IR divergence problem for QED \cite{yennie61,weinberg65}, which are now standard in many textbooks \cite{weinbergQFT}.

An opposing point of view argues instead that this information loss is illusory - that the IR modes are ``carried along'' with the relevant matter field \cite{gabai16,mirbab16,bousso17}. This point of view goes back to Chung \cite{chung65} and Kibble \cite{kibble68} (see also Kulish and Faddeev \cite{faddeevK70}), who argued that any calculation of the IR properties should be formulated in terms of coherent states for the background radiation field, in which no IR cutoff should be involved. According to this point of view, we do not average over the very low-energy bosons when trying to describe any information loss, and in fact there is no information loss (one can however formulate a contrary point of view, also using coherent states \cite{carney1710}, see also \cite{ChoiAkhoury} for yet another perspective).

In order to address this question - about what is the information loss - we actually begin here by formulating a more general question, and show how to answer it. The question we address instead is: how can one describe and quantify the decoherence in a gravitational system, and what is the correct way to describe the information loss? The results we find are applicable way beyond the scattering problem - using a decoherence functional one can discuss any kind of information loss in the system, whether one deals with scattering or some quite different set-up.

In this paper we will also argue that a correct answer to this question requires a non-perturbative formulation, and moreover one which does not rely on either IR cutoffs introduced by hand, or on some set of putative measuring systems acting at future infinity. Thus a second question asks - how can one formulate the problem of information loss non-perturbatively?

To deal with these various questions we introduce two new techniques in this paper, viz.,

(i) we introduce a ``composite generating functional" which, amongst other things, allows us to calculate the time evolution of the matter field reduced density matrix. This generating functional is a generalization of the decoherence functional well known in condensed matter physics \cite{decoF}; in the present case we specialize to the case of a scalar matter field coupled to gravitons.

(ii) To formulate the infrared physics non-perturbatively, we adapt a technique originally devised by Fradkin and collaborators \cite{fradkin}, which takes the form of a WKB expansion about the eikonal limit. No coherent boson states or IR cut-offs are required in this formulation. This allows us to directly address the controversy, discussed from different points of view in refs. \cite{stromSG,hawking16,hawking17,HPY14,ACS14,strom14,laddha15}
as well as refs. \cite{stromN,stromP,gabai16,mirbab16,bousso17,carney17,carney1710,ChoiAkhoury}, over information loss in graviton scattering.

Using these techniques we derive a functional eikonal expansion for the composite generating functional of a scalar field interacting with the gravitational field, written in terms of pairs of Feynman paths $T_{\mu\nu}(x), T^{\prime}_{\mu\nu}(x)$ for the matter field stress energy - this is our principal new result. We then look at the scattering problem that has caused so much discussion. To do this one needs to further extend the composite generating functional technique, to calculate the scattering of a reduced density matrix for the matter field and its ``in" and ``out" states. We then discover that in the asymptotic limit where these states are very widely separated, the decoherence functional can be written in terms of the asymptotic Bondi-Metzner-Sachs (BMS) charges for the system, as well as in terms of a gravitational memory function. In this way we confirm that the information loss can be written in terms of these charges, as argued by Strominger and others \cite{stromSG,hawking16,hawking17,HPY14,ACS14,strom14,laddha15,stromN}.

The rough plan of this paper is as follows. In the next section (section 2) we describe the basic formalism used in this work. We introduce the composite propagators and generating functional used here, giving detailed expressions for a scalar matter field coupled to gravitons; and we then show how these can be used to derive the decoherence functional for the matter field. We also give a brief discussion of how one deals with diffeomorphism invariance in this formalism. Then, in section 3, we describe the eikonal expansion technique used here to isolate out the key infrared (IR) behaviour that we are interested in - this involves first making a formal separation between slow and fast variables, and then making a functional eikonal expansion for the graviton variables, to give expressions for quantities like the decoherence functional introduced in section 2.

In sections 4 and 5 we move on to discuss the scattering problem for the matter field. We first derive general results for the ``composite $S$-matrix" of the reduced density matrix (this is {\it not} the scattering matrix for the fields themselves), in terms of our composite propagators, and show how this can be written in terms of the decoherence functional $\Gamma[T,T']$. Finally, in section 5, we show how both of these can be written as a function of the BMS Noether charges and in terms of a gravitational memory function; and we summarize the extent to which these results answer the questions posed in this introduction.

In this paper we will assume that readers are familiar with the relevant techniques in relativistic field theory and quantum gravity, but not necessarily with decoherence functionals, and so we give some introductory explanation of these.


\section{II: Composite Generating Functional and Influence Functional}
\label{sec:CGenF-SG}


In this section we introduce the formal tools to be used, as well as establishing our notation. In particular, we

(i) describe ``composite propagators" and the associated composite generating functionals. To make this clear we do it both for ordinary quantum mechanics, and for a matter field coupled to soft gravitons, after integrating out the soft gravitons.

(ii) derive the form of the Feynman influence functional for the matter field. For those unfamiliar with influence functionals and the related decoherence functional, we give a short introduction to these.

We also add brief remarks on the diffeomorphism invariance of the techniques used.

\subsection{II.1: Composite Generating Functionals}
\label{sec:CGF}

Suppose that in quantum field theory we find it necessary to consider a set of ``conditional propagators'', where propagators are defined between initial and final states, subject to the restriction that the values of certain functionals of the paths must take certain specified values at specific points in spacetime. An equivalent problem would be to define ``conditional correlation functions", such that one measures some correlator with the added restriction that the system propagates between a specified set of intermediate field values, which may also include specified initial and final states.

Such objects are no more than generalizations of functions sometimes considered in traditional QM, in the theory of measurement, wherein one calculates the expectation values of operators (including propagators) with the added specification that measurements of certain operators are performed, with specified results, during the propagation (see, eg., refs. \cite{ABL64,schmid87}). As an example, consider in ordinary non-relativistic QM the conditional propagator
\begin{equation}
\chi(x,x'| \{ A_{\alpha} \}) \;=\; \langle x| \hat{T} \{ \hat{A}_1(t_1), \cdots \hat{A}_p(t_p) \}|x' \rangle
 \label{chiA}
\end{equation}
which has the path integral representation
\begin{eqnarray}
\chi(x,x'| \{ A_{\alpha} \}) \;&=&\; \int {\cal D}p(t) \int_{x'}^{x} {\cal D} q(t) \; e^{{i \over \hbar} \int dt (p \dot{q} - {\cal H}(q,p))} \nonumber \\    && \qquad \times  \prod_{\alpha = 1}^p A_{\alpha}(p(t_{\alpha}), q(t_{\alpha}))
 \label{chiA-PI})
\end{eqnarray}
where $p$ and $q$ are canonically conjugate coordinates, with the operators $\hat{A} \equiv A(\hat{p}, \hat{q})$ replaced by functions $A(p, q)$ in the path integral, and where ${\cal H}(p,q)$ is the Hamiltonian. Physically we can interpret this either as (i) the amplitude to propagate from $x'$ to $x$, with the stipulation that at intermediate times, measurements of the operators $\hat{A}_{\alpha}$ at times $t_{\alpha}$ have given results $A_{\alpha}$, or (ii) the correlation between measurements of the operators $\hat{A}_{\alpha}$ at times $t_{\alpha}$, with the stipulation that the system starts at $x'$ and finishes at $x$.

In what follows we set up a way of computing objects of this kind in quantum gravity. We begin by recalling the usual definitions for (i) propagators, and (ii) correlation functions, in quantum gravity; we then define the conditional propagators.   This will be done in a Keldysh/Schwinger formalism, both for the generating functional\cite{SdW-GF,schmid82}, and for propagators and correlation functions.

\subsubsection{II.1 (a) Propagators and Correlators}
 \label{sec:propcorr}

The following material is standard in quantum field theory \cite{weinbergQFT}; we simply establish our notation here. We define ordinary propagators in the usual way as path integrals, so that, eg., a single particle has the propagator
\begin{eqnarray}
K_2(x,x') \;&=&\;  \int {\cal D}g^{\mu \nu} \Delta(g) \; e^{{i \over \hbar} S_G[g^{\mu \nu}]} \int_{x'}^x {\cal D}q \; e^{{i \over  \hbar}  \; S_o[q, g^{\mu \nu}]} \nonumber \\
\;&=&\; \int {\cal D}g^{\mu \nu} \Delta(g) \; e^{{i \over \hbar} S_G[g^{\mu \nu}]} \; K_2(x,x'|g)
 \label{K2-QM-int-g}
\end{eqnarray}
where $K_2(x,x'|g)$ is the propagator in a fixed background metric $g^{\mu\nu}(x)$, ie.,
\begin{eqnarray}
K_2(x,x'|g) &\;\;=\;\;&   \int_{x'}^x {\cal D}q \; e^{{i \over \hbar} S_o[q, g^{\mu\nu}]}  \nonumber \\
 & \;\; \equiv \;\; & \int_{x'}^x {\cal D} q(s) e^{-{i \over \hbar} \int ds [{m \over 2} g_{\mu \nu}(q) \dot{q}^{\mu} \dot{q}^{\nu} ]} \;\;\;\;\;\;
 \label{K2'-QM}
\end{eqnarray}
and $S_G[g^{\mu\nu}]$ is the Einstein action; in (\ref{K2'-QM}) we have written $dq^{\mu}/ds = \dot{q}^{\mu}(s)$.

In the same way, for a field $\phi(x)$ which propagates between configurations $\Phi'(x)$ and $\Phi(x)$, in a fixed background metric $g^{\mu \nu}(x)$, we have
\begin{eqnarray}
K_2(\Phi,\Phi') \;&=&\;  \int {\cal D}g^{\mu \nu} \Delta(g) \; e^{{i \over \hbar} S_G[g^{\mu \nu}]} \int_{\Phi'}^{\Phi} {\cal D}\phi \; e^{{i \over  \hbar}  \; S_M[\phi, g^{\mu \nu}]} \nonumber \\ \;&=&\; \int {\cal D}g^{\mu \nu} \Delta(g) \; e^{{i \over \hbar} S_G[g^{\mu \nu}]} \; K_2(\Phi,\Phi'|g)
 \label{K2-QFT-int-g}
\end{eqnarray}
where $\Delta(g)$ is a Faddeev-Popov determinant which divides out diffeomorphism-equivalent metric configurations, and $S_M[\phi, g^{\mu \nu}]$ is the scalar field action in the background $g^{\mu\nu}(x)$.

We also define $n$-point correlators in the usual way, as correlation between the field at $n$ different spacetime points, defined as functional derivatives of the partition function or generating functional. We define the generating functional using a Keldysh path integration which proceeds from a time slice at past infinity, out to future infinity, and back again \cite{SdW-GF,schmid82}. At finite temperature we will close this path at past infinity, with a ``ring" integration \cite{bloch} along an imaginary time continuation around a cylinder of circumference $1/kT$.

\begin{figure}
\includegraphics[width=3.2in]{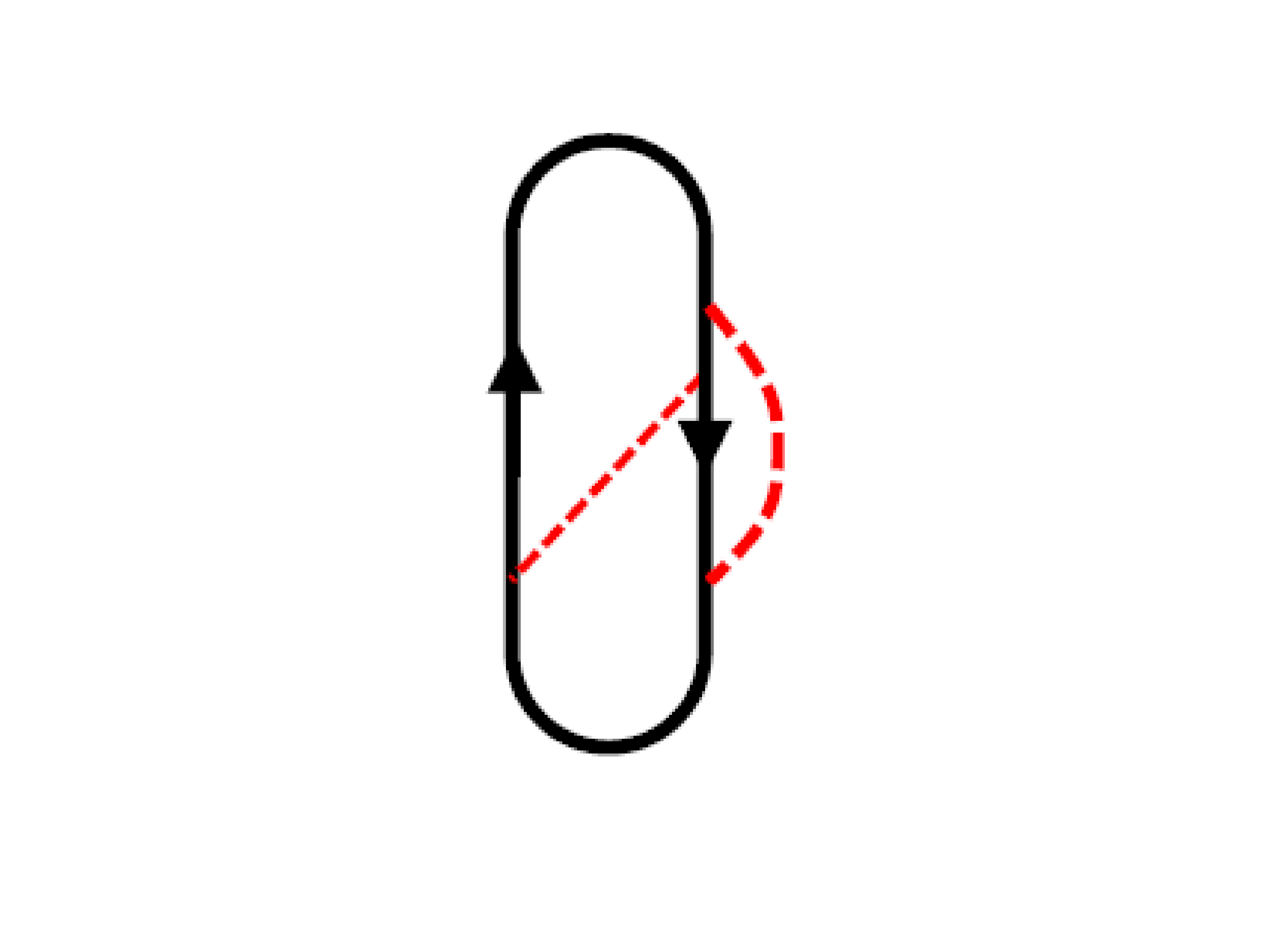}
\caption{\label{Fig:keldysh} A diagrammatic representation of the ``ring" Keldysh propagator for a particle coupled to gravitons; the thick line is the particle propagator, and the red hatched lines are gravitons. }
\end{figure}

Thus, eg., consider a relativistic single particle, moving in a fixed background metric, with the "ring" generating functional
\begin{eqnarray}
 {\cal Z}_o[j] &=& \oint {\cal D}g^{\mu \nu} \Delta(g) \; e^{{i \over \hbar} S_G[g^{\mu \nu}]} {\cal Z}_o[g;j]
 \label{Q0-jg}
\end{eqnarray}
where $j_{\mu}(x)$ is a `current' source function of proper time, coupling linearly to $q(s)$, and ${\cal Z}_o[g;j] = \oint {\cal D} q(s) e^{{i \over \hbar}(S_o[q;g] + \int jq)}$ is the particle partition function in a background metric $g^{\mu\nu}(x)$. The ring amplitude is defined over a Keldysh path, with the Keldysh integration denoted by $\oint$; see Fig. \ref{Fig:keldysh} for a graphical depiction of this.


The correlation functions $G_n^{\sigma_1,..\sigma_n}(s_1,..s_n)$ are then defined in the usual way by the functional differentials of ${\mathcal{Z}}_o[j]$ at proper times $s_j$, with $j = 1,2,..n$, where the index $\sigma_j = \pm$ indicates which section of the loop (forward or backward) the functional differential is being taken:
\begin{equation}
G_n^{\sigma_1,..\sigma_n}(s_1,..s_n) \;=\;  \left( {-i\hbar} \right)^n  \lim_{ j(s) \rightarrow 0} \left[ { \delta^n {\cal Z}_o[j] \over \delta j(s_1 \sigma_1) .. \delta j(s_n \sigma_n)} \right]
 \label{Gn-NR}
\end{equation}

\subsubsection{II.1 (b) Conditional Propagators}
 \label{sec:condprop}

The idea and use of conditional propagators is less common in quantum field theory, but follows naturally from the above. A simple example for a scalar field would be the conditional propagator in which the field $\phi(x)$ itself is measured at $p$ different spacetime points $\{ x_{\alpha} \}$, with $\alpha = 1, \cdots p$, during its propagation between initial and final configurations $\Phi'(x)$ and $\Phi(x)$; this is
\begin{widetext}
	
\begin{eqnarray}
\chi^{(p)}_2 (\Phi,\Phi'| \{ \phi(x_{\alpha}) \} )
\;&=&\; (-i \hbar)^p { \delta^p \over \delta J(x_1) ....\delta J(x_p)} \; K_2(\Phi,\Phi'|J(x)) \; \Big|_{J=0} \nonumber \\
&=&\; (-i \hbar)^p { \delta^p \over \delta J(x_1) ....\delta J(x_p)} \int {\cal D}g^{\mu \nu} \Delta(g) \; e^{{i \over \hbar} S_G[g^{\mu \nu}]} \int_{\Phi'}^{\Phi} {\cal D} \phi \; e^{ {i \over \hbar}( S_M[\phi;g] + \int d^4x J(x) \phi(x) ) } \; \Big|_{J=0} \;\;\;\;\;\;\;\;\;\;
 \label{chi-p-QM}
\end{eqnarray}
with $p$ different external current insertions $J(x_{\alpha})$, at spacetime points $x_{\alpha}$, during the propagation of $\phi(x)$ between the initial and final configurations.

This result is the simplest of its kind. More generally we are interested in the analogue of (\ref{chiA-PI}) for the conditional propagator, involving conditions on the results of measurements of field operators. Considering again the example of scalar fields, we must then look at excitations of the vacuum state of form $|j\rangle\equiv\mathcal{O}_{j}(\phi)|0\rangle$; an example would be a simple product of field operators $\mathcal{O}_{j}(\phi)=\phi(x_{j_{n}})\cdots\phi(x_{j_{1}})$. We will then imagine field operators $\hat{\cal A}_{\alpha} (\phi)$ able to act on the scalar fields on any intermediate time slice between initial and final field configurations $|1\rangle = \mathcal{O}_{1}(\phi)|0\rangle$ and
$|2\rangle = \mathcal{O}_{2}(\phi)|0\rangle$. The composite propagator $\chi^{(p)}_2 (2, 1| \{ {\cal A}_{\alpha} (\phi) \} )  \; \equiv \; \chi^{(p)}_2 (\mathcal{O}_{2}, \mathcal{O}_{1}| \{ {\cal A}_{\alpha} (\phi) \} )$ which describes this process is then

\begin{equation}
\chi^{(p)}_2 (2, 1| \{ {\cal A}_{\alpha} (\phi) \} )
\;=\; \prod_{\alpha = 1}^p {\cal A}_{\alpha} \left({-i \hbar \delta \over \delta J(x_{\alpha})} \right) \; K_2(2,1|J(x)) \; \Big|_{J=0}
 \label{chi-A-phi}
\end{equation}
where the propagator $K_2(2,1|J(x)) \; \equiv \; K_2(\mathcal{O}_{2}, \mathcal{O}_{1}|J(x))$ is written as
\begin{equation}
 K_2(2, 1|J(x)) \;=\; \mathcal{O}_{2}
\left({-i \hbar \delta \over \delta J(x)} \right)\,\mathcal{O}_{1}\left({-i \hbar \delta \over \delta J(x)} \right) \oint {\cal D}g^{\mu \nu} \Delta(g) \; e^{{i \over \hbar} S_G[g^{\mu \nu}]} \oint {\cal D} \phi \; e^{ {i \over \hbar}( S_M[\phi;g] + \int d^4x J(x) \phi(x) ) }
 \label{K-12-J}
\end{equation}
\end{widetext}
with the path integrals over Keldysh contours.

In what follows we will always write composite functionals for quantum fields as path integrals over Keldysh contours, and we will write the gravitational path integral as a graviton expansion around a flat spacetime. The gravitational action is specified by writing $g^{\mu \nu} = \eta^{\mu\nu} + 2\kappa h^{\mu\nu}$ in the usual way, with $\kappa^2 = M_P^{-2} = 8\pi G$. We can then write the action in the form
\begin{equation}
{\cal S} \;=\; {\cal S}_{\phi} + {\cal S}_G + {\cal S}_{int}
 \label{Stot}
\end{equation}
in which ${\cal S}_{\phi} \; =\; \int d^4x \; \mathcal{L}_{\phi}$ is the matter Lagrangian, here that of a scalar field, where the graviton action is
\begin{equation}
{\cal S}_G \;=\; {\cal S}_{YGH} - \int d^4x h^{\mu\nu}(x) \bar{G}_{\mu\nu}(x)
 \label{SG}
\end{equation}
in which $\bar{G}_{\mu\nu}$ is the linearized Einstein tensor, viz.,
\begin{align}
\bar{G}_{\mu\nu}=\frac{1}{2}\big(&-\partial^{2}h_{\mu\nu}-\partial_{\mu}\partial_{\nu}h+\partial^{\rho}\partial_{\mu}h_{\rho\nu} \nonumber \\
& +\partial^{\rho}\partial_{\nu}h_{\rho\mu} -\eta_{\mu\nu}\partial^{\sigma}\partial^{\rho}h_{\sigma\rho}+\eta_{\mu\nu}\partial^{2}h\big)
\end{align}
and ${\cal S}_{YGH}$ is the linearized York/Gibbons-Hawking term \cite{york72}; and the matter-gravity coupling term is
\begin{equation}
{\cal S}_{int} \;=\; \kappa \int d^4x h^{\mu\nu}(x) T_{\mu\nu}(x)
 \label{Sint}
\end{equation}
in which  $T^{\mu\nu}(x) \equiv T^{\mu\nu}(\phi(x))$ is the matter stress-energy tensor, viz.,
\begin{equation}
T_{\mu\nu}=-\partial_{\mu}\phi\partial_{\nu}\phi-\eta_{\mu\nu}\mathcal{L}_{\phi}.
 \label{T-phi}
\end{equation}

\subsection{II.2: Density Matrix Dynamics}
\label{sec:CGF2}

We will be interested primarily in the direct calculation of probabilities for the matter field. We will thus be calculating reduced density matrices for the matter field, taken between two matter field configurations, having already integrated out the gravitons in a way which needs to be specified.

The dynamics of the reduced density matrix is written in terms of a propagator ${\cal K}(2,2'; 1,1')$ for the matter density matrix in the form
\begin{equation}
\rho_{\phi}(2,2')\;=\; \int d1\int d1' \mathcal{K}(2,2';1,1')\; \rho_{\phi}(1,1')
 \label{rho+K}
\end{equation}
where here the labels $1,1'$, and $2,2'$, refer to initial and final values respectively of the scalar matter fields $\phi(x), \phi'(x)$. These states will be specified as above, so that the states $|1 \rangle$, $|1' \rangle$, $|2 \rangle$, $|2' \rangle$ are introduced by inserting the relevant operators, so that, eg.,
\begin{equation}
|1\rangle \; \equiv \; \mathcal{O}_{1}(\phi)|0\rangle
 \label{states}
\end{equation}
and so on; thus $\rho^{\phi}(1,1')  \equiv \rho^{\phi}(\mathcal{O}_{1}(\phi), \mathcal{O}_{1'}(\phi))$. Formally, one always requires a pair of fields to define the evolution of a density matrix, referring to the ``forward" and ``backward" paths of the Schwinger/Keldysh propagator.

It will be useful in what follows to generalize the expression (\ref{rho+K}) to include a pair of external fields $J(x), J'(x)$ coupled to the scalar matter fields $\phi(x), \phi'(x)$ respectively. The goal is to give path integral expressions for the resulting propagator $\mathcal{K}(2,2';1,1'|J,J')$, in terms of a ``composite generating functional", which can itself be written in terms of a functional first defined by Feynman \cite{feynV63}, and known as the Feynman influence functional. We begin by giving formal expressions, and then explain their physical meaning \cite{calzetta}.

\subsubsection{II.2 (a) Density Matrix Propagator}
 \label{sec:formal}

Let's start by just listing the main formal results we will require. The idea is to begin with a density matrix for the total ``universe" (here this is the scalar matter field coupled to the gravitational field) and then trace over the gravitational environment to get the reduced density matrix for the matter field, so that
\begin{equation}
\hat{\rho}_{\phi} \;=\; Tr_G \; \mbox{\boldmath $\hat{\rho}$}_U
 \label{rhoR}
\end{equation}
We will assume that in the distant past the universal density matrix begins in an uncorrelated product form, viz.,
\begin{equation}
\mbox{\boldmath $\hat{\rho}$}_U^{(in)} \;=\; \hat{\rho}_{\phi}^{(in)} \bigotimes \hat{\rho}_G^{(in)}
 \label{rhoT}
\end{equation}
and that the gravitational density matrix $\hat{\rho}_G^{(in)}$ in the distant past can be described, in linearized gravity, by a thermal density matrix; the matter state is initially a vacuum state for the matter field. Entanglement between the matter field $\phi(x)$ and the gravitons is then generated by the gravitational coupling. The assumption of an initial product state is typically made to simplify the formal development.

One can write a path integral expression for the propagator $\mathcal{K}(2,2';1,1'|J,J')$ of the reduced density matrix as
\begin{widetext}
\begin{equation}
\mathcal{K}(2,2'; 1,1'|J, J')\;=\;\int \mathcal{D}\phi  \int \mathcal{D}\phi'
 \,\mathcal{O}_{2}(\phi)\, \mathcal{O}_{2'}(\phi')
\mathcal{O}_{1}(\phi) \mathcal{O}_{1'}(\phi') \; e^{i \left[ S_{\phi}[\phi] - S_{\phi}[\phi'] +\int d^{4}x\,(J(x)\phi(x) - J'(x)\phi'(x)) \right] }
\mathcal{F}[\phi, \phi']
 \label{K2211}
\end{equation}
where $\mathcal{F}[\phi, \phi']$ is the Feynman influence functional, defined below. By the standard manouevre of transforming from the fields $\phi, \phi'$ to their conjugate current variables $J(x), J'(x)$, we can also write (\ref{K2211}) in the form
\begin{equation}
\mathcal{K}(2,2'; 1,1'|J, J')\;=\;
 \,\mathcal{O}_{2}
\left({-i \hbar \delta \over \delta J(x)} \right)\,\mathcal{O}_{2'}\left({i \hbar \delta \over \delta J'(x)} \right)  \mathcal{O}_{1}
\left({-i \hbar \delta \over \delta J(x)} \right)\,\mathcal{O}_{1'}\left({i \hbar \delta \over \delta J'(x)} \right)
\mathscr{Z}[J(x), J'(x)]
 \label{K2211b}
\end{equation}
where the composite generating functional $\mathscr{Z}[J(x), J'(x)]$ is defined as
\begin{equation}
 \label{eq:compgenF}
\mathscr{Z}[J,J'] \;=\; \int \mathcal{D} \phi(x) \int \mathcal{D} \phi'(x) \; e^{i \left[ S_{\phi}[\phi] - S_{\phi}[\phi'] +\int d^{4}x\,(J(x)\phi(x) - J'(x)\phi'(x)) \right] }
\mathcal{F}[\phi, \phi']
\end{equation}
\end{widetext}


\begin{figure}
\includegraphics[width=3.2in]{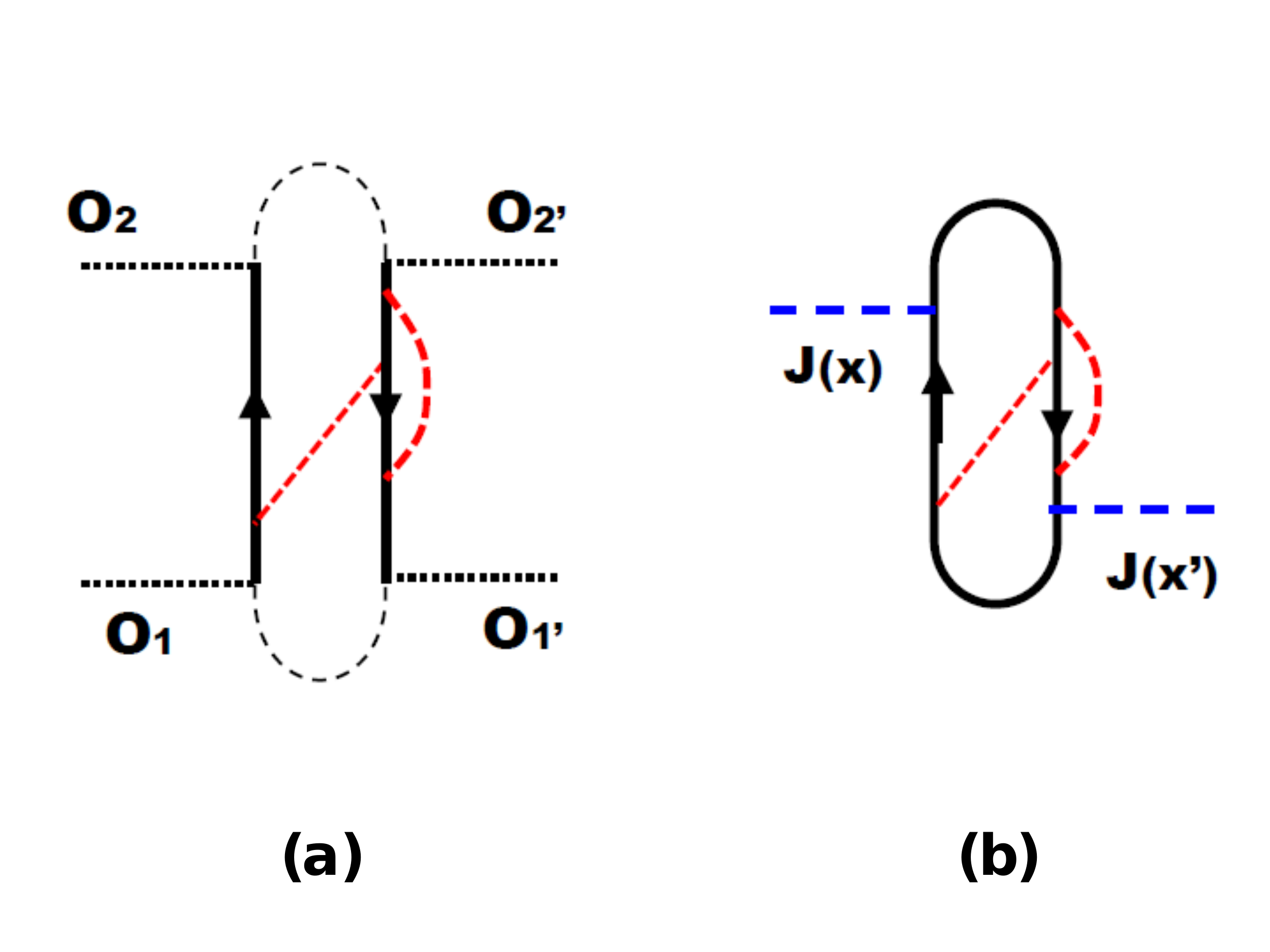}
\caption{\label{Fig:K21+ZJ} Graphical representation of typical terms in (a) the propagator for the matter field density matrix $\mathcal{K}(2,2'; 1,1')$, in the absence of any external currents, and (b) the composite generating functional $\mathscr{Z}[{\bf J}]$ in the presence of external currents $J(x), J'(x)$. The matter fields are shown in heavy black, the graviton propagators in hatched red, and external currents in hatched blue; the effect of the field operators $\mathcal{O}_{\bf j}$ is shown in finely hatched black. }
\end{figure}


These explicit expressions are rather lengthy, and it is useful to introduce here a compact notation for the Keldysh paths involved \cite{schmid82}, in which spacetime coordinates, fields and currents, etc., are all represented as 2-component boldface vectors, referring to the forward and backward segments of the paths. Then the equation of motion (\ref{rho+K}) for the matter field reduced density matrix $\rho^{\phi}$ becomes
\begin{equation}
\rho_{\phi}({\bf 2})\;=\; \sum_{{\bf 1}} \mathcal{K}({\bf 2};{\bf 1})\; \rho_{\phi}({\bf 1})
 \label{rho+K'}
\end{equation}
so that $\rho_{\phi}({\bf 1}) \equiv \rho(1,1')$ and $\mathcal{K}({\bf 2};{\bf 1}) \equiv \mathcal{K}(2,2'; 1,1')$. The result (\ref{K2211}) for the density matrix propagator is then written as
\begin{equation} \label{eq:dmpropagator}
\mathcal{K}({\bf 2};{\bf 1}|{\bf J})=\int \mathcal{D}
\mbox{\boldmath $\phi$} \,\mathcal{O}_{\bf 2}\,
\mathcal{O}_{\bf 1}e^{i [S_{\phi}[\mbox{\boldmath $\Phi$}] +\int d^{4}x\, {\bf J}\cdot \mbox{\boldmath $\Phi$}] }
\mathcal{F}[\mbox{\boldmath $\Phi$}],
\end{equation}
where the external source field ${\bf J} \equiv (J,J')$,
where $\mbox{\boldmath $\Phi$} \equiv (\phi, \phi')$, where $S_{\phi}[\mbox{\boldmath $\Phi$}] \equiv S_{\phi}[\phi] - S_{\phi}[\phi']$ and ${\bf J}\cdot \mbox{\boldmath $\Phi$} \equiv (J\phi - J'\phi')$, and where $\mathcal{F}[\mbox{\boldmath $\Phi$}]$ is the Feynman influence functional. The equivalent result (\ref{K2211}) for the density matrix propagator is written as
\begin{equation}
\mathcal{K}({\bf 2},{\bf 1})=\mathcal{O}_{\bf 2}
(\delta_{\bf J})\,\mathcal{O}_{\bf 1}(\delta_{\bf J})\mathscr{Z}[{\bf J}]\bigg|_{{\bf J}=0}.
 \label{ooz}
\end{equation}
and the composite generating functional is just
\begin{equation}
 \label{eq:DMgeneratingfunctional}
\mathscr{Z}[{\bf J}]=\int \mathcal{D} \mbox{\boldmath $\Phi$}  \,e^{i [S_{\phi}[\mbox{\boldmath $\Phi$}]+\int d^{4}x\, {\bf J}\cdot \mbox{\boldmath $\Phi$}]}\,\mathcal{F}[\mbox{\boldmath $\Phi$}].
\end{equation}
where $\delta_{\bf J} \equiv -i \hbar(\delta/\delta J(x), \; -\delta/\delta J'(x))$. One can give a graphical interpretation of the function $\mathcal{K}(2,2'; 1,1')$ appearing in eqns. (\ref{K2211}) and (\ref{ooz}) as shown in Fig. \ref{Fig:K21+ZJ}(a); the equivalent graphical interpretation of the composite generating functional $\mathscr{Z}[{\bf J}]$ is shown in Fig. \ref{Fig:K21+ZJ}(b).

We can see, by comparing either eqtn. (\ref{ooz}) or (\ref{K2211b}) with (\ref{K-12-J}), that $\mathscr{Z}[{\bf J}]$ is acting as an analogue of the usual generating functional in ordinary quantum field theory, but now for ``reduced" quantities like the reduced density matrix, in which the gravitons have already been integrated out. This is clear from the graphical representation in Fig. \ref{Fig:K21+ZJ}(b), in which functional integration over two graviton lines is shown.

\subsubsection{II.2 (b) Influence Functional}
 \label{sec:formal2}

All the key physics in our problem is in the influence functional
$\mathcal{F}[\phi, \phi']$; it not only describes the dephasing and relaxation of the matter field by the gravitational field, but also all reactive renormalization effects of gravitational interactions on the matter field.

Formally the influence functional is produced by integrating out the graviton and interaction terms (\ref{SG}) and (\ref{Sint}) in the density matrix propagator, so that we have
\begin{equation}
{\cal F}[{\bf T}] \;=\; \int {\cal D} {\bf h} \; e^{{i \over \hbar} (S_G[{\bf h}] + S_{\textrm{int}}[{\bf h}, {\bf T}])}
 \label{F-Th}
\end{equation}
where we are again using our compact notation, and we have written ${\cal F}$ as a functional of ${\bf T}$ instead of $\mbox{\boldmath $\Phi$}$, using (\ref{T-phi}).

It is convenient to write
\begin{equation}
{\cal F} = e^{i(\Psi_o + \Psi)},
 \label{F-PhiPsi}
\end{equation}
where $\Psi_o$ incorporates all static ``self-gravity" effects (the analogue of the Coulomb contribution in a QED calculation), and where the complex phase functional $\Psi[T, T']$ contains all
dynamic effects. In a linearized gravity theory, where the graviton action is a quadratic function of the field $h^{\mu\nu}(x)$, and the coupling a linear function, then the form of $\mathcal{F}[T,T']$ is trivially recovered as the exponential of a
quadratic form over the fields \cite{feynV63}. Separating the real and imaginary parts as $\Psi[T,T'] =  \Delta[T,T'] + i \Gamma[T,T']$ one has the explicit expressions  \cite{oniga16}:
\begin{widetext}
\begin{align}
 \label{eq:QGIF}
\Delta[T,T']\;\; =& \;\; {\kappa^2 \over 2}\int^{t_{f}}_{t_{i}}d^{4}x \int^{x^{0}}_{t_{i}}d^{4}\tilde{x} \left[T_{\mu\nu}(x)-T^{\prime}_{\mu\nu}(x) \right] D_1^{\mu\nu\alpha\beta}(x,\tilde{x}) \left[T_{\alpha\beta}(\tilde{x})+T^{\prime}_{\alpha\beta}(\tilde{x}) \right]  \nonumber \\
\Gamma[T,T']\;\; =&\;\; {\kappa^2 \over 2}\int^{t_{f}}_{t_{i}}d^{4}x \int^{x^{0}}_{t_{i}}d^{4}\tilde{x} \left[T_{\mu\nu}(x)-T^{\prime}_{\mu\nu}(x) \right] D_2^{\mu\nu\alpha\beta}(x,\tilde{x}) \left[T_{\alpha\beta}(\tilde{x})-T^{\prime}_{\alpha\beta}(\tilde{x}) \right]
\end{align}
where $D^{\mu\nu\alpha\beta}(x) = D_1^{\mu\nu\alpha\beta}(x) + iD_2^{\mu\nu\alpha\beta}(x)$ is just the finite-temperature graviton propagator:
\begin{equation}
D^{\mu\nu\alpha\beta}(x)\;=\; \int\frac{d^{3}q}{(2\pi)^{3}} {e^{i\mathbf{q}\cdot\mathbf{x}} \over q} \; \Pi^{\mu\nu\alpha\beta}(q)\left(\sin qx^0 + i\cos q x^{0} \coth {\beta q^0 \over 2}\right)
 \label{Dgrav}
\end{equation}
defined at temperature $T$, where we write $\beta = 1/kT$, and where $\Pi_{\mu\nu\alpha\beta}(q)$ is the ``TT projector", which projects out all but the transverse traceless modes. Note that (\ref{eq:QGIF}) is for the moment formal, since we have yet to specify how to deal with high-energy cutoffs, etc.
\end{widetext}

The imaginary part $\Gamma[T,T']$ of the influence functional phase is what is usually referred to as the ``decoherence functional" \cite{decoF}, and is of primary interest to us here. Once exponentiated and inserted into the composite generating functional, its physical meaning is most easily understood by expanding the exponential. A 4th-order (in $kappa$) term is shown in Fig. \ref{Fig:K21+ZJ}; we see that it generates both ``self-energy" graviton interactions on one or other of the matter lines, or an interaction between the forward and return lines.

The result of the interactions between lines is to cause dephasing in the dynamics of the matter field - this can happen even at $T=0$, if accelerations are involved in the dynamics of the matter field - this will then lead to the emission of soft gravitons. At finite $T$, the matter field is interacting with a thermal bath, which has a well-defined rest frame - in this case  one also has real relaxation processes caused by inelastic scattering of the gravitons.

\subsubsection{II.2 (c): Questions of Gauge Invariance}
\label{sec:gauge}

Let us briefly comment here on the use of the TT projector in eq. (\ref{eq:QGIF}); see also ref. \cite{wilson17}. The need to satisfy both constraints and gauge/diffeomorphism invariance persists even in linearized gravity. In linearized gravity  $h_{\mu\nu}$ is treated as a dynamical variable; however, not all components are independent.  By linearizing, we break the full diffeomorphism invariance of GR; nevertheless small diffeomorphisms, for which $\kappa h_{\mu\nu}\ll 1$, are still gauge symmetries of linearized gravity, so that not all components of $h_{\mu\nu}$ are physical. Likewise, in a Hamiltonian formalism not all components of $h_{\mu\nu}$ are independent canonical coordinates. The timelike components $h_{0\nu}$ do not have conjugate momenta since $\pi^{0\nu}=\partial \mathcal{L}/\partial(\partial_{0}h_{0\nu})$ vanishes identically; and the timelike components of the linearized Einstein equation are not dynamical equations describing the time evolution of phase space data $(h_{jk},\pi^{jk})$, since they involve no time derivatives of the canonical variables.  Instead, the timelike linearized Einstein equations impose constraints on the phase space data which restrict what configurations can even exist on a time-slice.

In that subspace of phase space where the constraints are satisfied only the transverse-traceless components are independent. It is trivial to check that these components of the metric are invariant under gauge transformations; hence, if the constraints of linearized gravity are satisfied, then equations written in terms of the remaining variables will be gauge invariant. This is true in any quantum theory if the constraints are treated via Dirac quantization~\cite{oniga16, DiracConstraints}. This is why the transverse-traceless projectors appear in the interaction kernels in eq.~\eqref{eq:QGIF} - they project the influence functional onto the appropriate constrained subspace.

\subsubsection{II.2 (d): Summary}
\label{sec:summary2}

It is helpful here to summarize our basic approach. In quantum field theory one typically starts from the generating functional of correlation functions, from which various transition amplitudes are obtained via functional differentiation. The evolution of the full system is unitary and described by the standard path integral.

Here we have introduced an analogous object for an open quantum system, the composite generating functional in eq.~\eqref{eq:DMgeneratingfunctional}. We can then study the evolution of generic matter density matrices coupled to an unobserved background of gravitons in a way entirely parallel to typical quantum field theory computations, by taking functional derivatives with respect to external currents. This new formalism contains at its heart a decoherence functional, which describes both phase decoherence and relaxation processes.

Clearly one should be able to extend this formalism to cover scattering processes, involving multi-particle ``in" and ``out" states, (in a way analogous to that LSZ formalism of standard quantum field theory).  We develop these ideas in sections 4 and 5 below.


\section{III: Functional Eikonal Expansion}
\label{sec:eikon}


As just noted, in sections 4 and 5 we will be applying our formalism to attack a problem of current interest, viz., information loss in graviton scattering processes. However before doing so we must make a detour, because we require a method which can deal in a fully non-perturbative way with soft gravitons. A standard Feynman diagrammatic perturbation approach is not well suited to this problem, because the graviton is a massless particle, so that perturbation theory is plagued with infrared divergences

In this section we will use a more appropriate non-perturbative treatment of the path-integral, a functional eikonal expansion, which is well suited to situations in which there is a separation of scales - in the present case provided by massive particles coupled to long wavelength gravitons. This will also allow us to give a meaning to the more formal expressions in the last section.

In what follows we begin by discussing how one makes a formal separation of scales, and then give the functional eikonal expansion for the matter propagator and for the composite generating functional.

\subsection{III.1: Separation of Slow and Fast variables}
\label{sec:fastSlow}

The first thing we wish to do is make a formal separation between fast and slow variables in the composite functionals introduced above, in order to isolate out the interesting infrared behaviour. To do this we introduce a cutoff scale $\Lambda_0$ separating ``soft" gravitons (with momentum $|q|\leq\Lambda_{0}$) from  ``hard'' gravitons $(|q|>\Lambda_{0})$. In the course of our calculation we'll restrict the value of $\Lambda_{0}$, so that $\Lambda_0 \ll$ scalar particle masses.

Now since the interaction kernels $D_1$ and $D_2$ in the decoherence functional (\ref{eq:QGIF}) are given by a sum over contributions from each mode, the influence functional can be factored into hard and soft parts, ie., we can write
\begin{equation}
\mathcal{F}[\bm{\Phi}] \;=\;\mathcal{F}_{S}[\bm{\Phi}]\mathcal{F}_{H}[\bm{\Phi}]
 \label{sepHS}
\end{equation}
where $\Lambda_{0}$ serves as a UV cut-off in $\mathcal{F}_{S}[\bm{\phi}]$ and an IR cut-off in $\mathcal{F}_{H}[\bm{\phi}]$. However, we cannot similarly factorize the composite generating functional - instead, one has
\begin{align}
\mathscr{Z}[\textbf{J}] \;&=\; \mathcal{F}_{H}[\delta_{\textbf{J}}]\int \mathcal{D}\bm{\Phi} \;\; e^{i (S_{\phi}[\bm{\Phi}]+\int d^{4}x\, {\bf J}\cdot \bm{\Phi})}\;\mathcal{F}_{S}[\bm{\Phi}] \nonumber \\
&\equiv \mathcal{F}_{H}[\delta_{\textbf{J}}]\mathscr{Z}_{S}[{\bf J}].
 \label{ZsFs}
\end{align}
where again we use the shorthand $\mathcal{F}_{H}[\delta_{\bf J}] \equiv \mathcal{F}_{H}[\bm{\Phi}\rightarrow-i\delta/\delta {\bf J}]$ introduced above in (\ref{ooz}); this transformation
pulls the hard influence functional ${\cal F}_H$ outside the path-integral as a functional differential operator. Thus there is no result like $\mathscr{Z} = \mathscr{Z}_H \mathscr{Z}_S$, and so we must keep the hard matter in the calculation, even while keeping only the soft gravitons.

Since ${\cal F}_H$ has an IR cutoff $\Lambda_0$, one can expand it perturbatively - we will study this series in future work. Here we will focus on the contributions to decoherence from soft gravitons, encapsulated in the soft composite generating functional $\mathscr{Z}_{S}[{\bf J}]$, which generates the propagators describing the evolution of the matter density matrix under the influence of soft gravitons.

We now rewrite the soft generating functional in a crucial way. Recall that it is always possible to rewrite a matter field propagator coupled to some dynamic field (here $h^{\mu\nu}(x)$) in the form of a propagator in some fixed or ``frozen" background field configuration, with a subsequent functional integration over these field configurations. Accordingly we do this for the influence functional itself, pulling it outside of the path integral as a functional differential operator, to get
\begin{eqnarray}
\mathscr{Z}_S[{\bf J}] &=& \mathcal{F}_{S}[\delta_{\bf h}]\int
\mathcal{D}\mbox{\boldmath $\Phi$}\,e^{i (S_{\phi}[\mbox{\boldmath $
\Phi$}]+\int \,(\frac{1}{2}\kappa {\bf T}^{\mu\nu}\cdot{\bf h}_{\mu\nu}+ {\bf J}\cdot \mbox{\boldmath $\Phi$}))}\bigg|_{{\bf h}=0}
\nonumber \\
& \equiv & \mathcal{F}_{S}[\delta_{\mathbf{h}}]\; Z[J|h] \; Z^{*}[J'|h']
\bigg|_{h,h'=0}
 \label{Z-ZJ}
\end{eqnarray}
where in the 2nd expression we write the forward and backward path variables explicitly. Here $\mathcal{F}_{S}[\delta_{\bf h}]$ is defined by the substitution  $\mathcal{F}_{S}[\delta_{\bf h}] \equiv \mathcal{F}_{S}[{\bf T}_{\mu
\nu}\rightarrow-2iM_{P}\delta/\delta {\bf h}^{\mu\nu}]$, ie., $T_{\mu\nu}$ and $T'_{\mu\nu}$ are substituted by their conjugate variables, and $Z[J;h]$ is the generating functional for a scalar field in the slowly-varying background metric perturbation $h^{\mu\nu}(x)$, ie., we have
\begin{equation}
Z[J|h] = \int \mathcal{D}\phi\,e^{iS_{0}[\phi]+i\int d^{4}x\,(\frac{1}{2}\kappa T^{\mu\nu}h_{\mu\nu}+J\phi)}
 \label{Z-Jh}
\end{equation}

We see that in (\ref{Z-ZJ}) we have decoupled the primed and unprimed variables, so that from (\ref{ooz}) we have the density matrix propagator
\begin{align}
\mathcal{K}_S({\bf 2},{\bf 1}) &= \mathcal{O}_{\bf 2}
(\delta_{\bf J})\,\mathcal{O}_{\bf 1}(\delta_{\bf J}) \; \mathcal{F}_{S}[\delta_{\mathbf{h}}]  \nonumber \\ & \qquad\qquad \qquad \times   Z[J|h] \; Z^{*}[J'|h']
\bigg|_{h,h',{\bf J}=0} \qquad
 \label{oofz}
\end{align}

Note that since the influence functional is a functional of only the transverse-traceless parts of the stress tensor, the auxiliary field $h_{\mu\nu}$ is also transverse-traceless in addition to being slowly-varying. We will use these properties many times throughout the following derivation.

We can now make the calculations much more physically concrete. We'll assume for now that aside from interactions with soft gravitons the scalar field is free, in which case
\begin{equation}
S_{0}[\phi]=-\frac{1}{2}\int d^{4}x\,\partial_{\mu}\phi\partial^{\mu}\phi+m^{2}\phi^{2},
\end{equation}
and
\begin{equation}
T_{\mu\nu} = -\partial_{\mu}\phi\partial_{\nu}\phi+\frac{1}{2}\eta_{\mu\nu}(\partial_{\lambda}\phi\partial^{\lambda}\phi+m^{2}\phi^{2}).
\end{equation}
The path integral for $Z[J|h]$ is then Gaussian and can be written, within a normalizing constant, as
\begin{eqnarray}
Z[J|h] &=& \; e^{\frac{1}{2}\sum_{n=1}^{\infty}\textrm{Tr}(\kappa G_{0} h_{\mu\nu}\hat{\tau}^{\mu\nu})^{n}} \nonumber \\
& \; &   \qquad \times \; e^{-\frac{i}{2}\int d^{4}xd^{4}y\, J(x)\mathcal{G}(x,y|h) J(y)} \;\;\;\;\;
 \label{Z2-Jh}
\end{eqnarray}
in which $\hat{\tau}_{\mu\nu}\equiv\partial_{\mu}\partial_{\nu}-\frac{1}{2}\eta_{\mu\nu}(\partial^{2}-m^{2})$ is a differential operator corresponding to the stress-energy density of a point-particle, and $\mathcal{G}(x,y|h)$ is the scalar Green function on a fixed background $h^{\mu\nu}(x)$, ie.,
\begin{equation}
 \label{eq:waveequationonbackground}
(\partial^{2}-m^{2}+\kappa h^{\mu\nu}\hat{\tau}_{\mu\nu})\,\mathcal{G}(x,y|h) \;=\; \delta^{4}(x-y).
\end{equation}
We have already, and will continue to, make use of the flat-space limiting form $G_{0}(x,y) \equiv \mathcal{G}(x,y|h = 0)$.

Notice that in Eq.~\eqref{Z2-Jh} the first exponential factor is just a rewriting of the functional determinant of the  differential operator in (\ref{eq:waveequationonbackground}). If one thinks in terms of Feynman diagrams, a term in the series involving the trace of $n$ factors of $G_{0}h_{\mu\nu}\hat{\tau}^{\mu\nu}$ corresponds to a closed scalar loop with $n$ insertions of the external field. These diagrams describe vacuum polarization effects and will be suppressed by powers of $\Lambda_o/m \ll 1$. In this limit the polarization diagrams are then negligible. We can therefore omit the first exponential factor in (\ref{Z2-Jh}) and use the simpler result
\begin{equation}
Z[J|h]=e^{-\frac{i}{2}\int d^{4}xd^{4}y\, J(x)\mathcal{G}(x,y|h)J(y)}
 \label{Z-JJG}
\end{equation}
This result for $Z[J|h]$, along with the result (\ref{oofz}) for the generating functional and our equation of motion for ${\cal G}(x,y|h)$, will then be the starting point for the formal eikonal expansion. Note that if we wish, we can also include non-gravitational interactions in this expression, by adding the usual $\exp(i\int \mathcal{L}_{\textrm{int}}[\delta_{J}])$ factor as a prefactor on the right-hand side~\cite{footnote}.

\subsection{III.2: Functional Eikonal expansion for gravitons}
\label{sec:eikonExp}

The key intuition underlying any eikonal expansion is one of scale separation. In the absence of a background $(h_{\mu\nu}=0)$ the free Green's function $G_{0}(x,y)$ describes a single relativistic scalar particle propagating between spacetime points $y$ and $x$. The dominant path is the straight-line classical solution, with corrections from quantum fluctuations about this path. In the presence of a slowly-varying background, the leading order eikonal approximation only modifies the propagator with an `eikonal phase' accumulated along the classical path. From a perturbative standpoint this technique is very powerful - even the leading term sums an infinite class of diagrams, which here will capture in a non-perturbative way the leading IR-divergent effects from soft gravitons. Higher corrections capture sub-dominant contributions.

There are various ways to set up an eikonal expansion; in this paper we do this by adapting functional methods first introduced by Fradkin and collaborators \cite{fradkin}.

\subsubsection{III.2 (a): Equation of Motion}
\label{sec:eikonEOM}

We begin by re-writing the equation of motion for $\mathcal{G}(x,y|h)$ in (\ref{eq:waveequationonbackground}) after a partial Fourier transform. To separate fast and slow gravitons we note that on a slowly-varying background the propagator is a rapidly varying function of the relative coordinate $(x-y)$ (on a scale $\sim 1/m$ but varies slowly with the ``center of mass'' coordinate $X\equiv(x+y)/2$.   We define the partial Fourier transform over $(x-y)$ according to
\begin{equation}
\mathcal{G}(x,y|h) = \sum_k e^{ik(x-y)}\mathcal{G}_k(X|h)
 \label{FT-GF}
\end{equation}
to move the fast modes into momentum space.

We then have the equation of motion
\be\label{eq:eom}
\left\{G^{-1}_{0}(k)-\hat{\mathcal{H}}_{h}(x,k;h)\right\} \mathcal{G}_k(X|h) = 1,
\ee
where the inverse of the free propagator is just
\begin{equation}
G_0^{-1}(x)=-\partial^{2}-m^{2}
 \label{G-0}
\end{equation}
so that $G_0(k) = -(k^2+m^2)$, and we've also introduced the ``Hamiltonian'' operator
\begin{eqnarray}
\hat{\cal H}_h(x,k;h) &=& -(\partial^2 - 2ik^\mu\partial_\mu) \; + \nonumber \\
 &\;\;\;\;\;\;& \kappa h^{\mu\nu}(x)\left[   k_\mu k_\nu + 2ik_\mu\partial_\nu - \partial_\mu \partial_\nu \right] \;\;\;
 \label{calH-FT}
\end{eqnarray}
The sense in which this differential operator is a Hamiltonian will soon become clear.

We now introduce the Schwinger/Fock ``proper time'' representation for the propagators. The bare propagator is written as
\begin{eqnarray}
G_0(k) \;&=&\; -i\int_0^{\infty} ds\;e^{-is (k^2 + m^2)} \nonumber \\
&\equiv&\;  -i \int_0^{\infty} ds \,  G_0(k,s),
\end{eqnarray}
where we haven't written explicitly the small $i\epsilon$ convergence factor in the exponent. The Schwinger parameterized propagator $G_0(k,s)$ satisfies the equation of motion
\be
i\partial_s G_0(k,s) = (k^2 + m^2) G_0(k,s),
 \label{eq:15}
\ee
subject to the initial condition $G_{0}(k,s)|_{s=0}=1$.
To separate the fast variables from the slow variables we assume an explicit factorization in the Schwinger parameterization of the full propagator
\begin{equation} \label{eq:schwingrep}
\mathcal{G}_k(X|h) = -i\int_0^{\infty} ds \,  G_0(k,s) \,   \mathcal{Y} (k,s,X|h),
\end{equation}
such that $\mathcal{Y}$ acts to weight the free propagator term under the proper time integral. This is completely analogous to the elementary application of the WKB approximation in which one assumes that the Schr\"{o}dinger equation with a slowly varying potential is solved by a plane wave with a slowly varying amplitude and wavelength. The equation of motion eq.~\eqref{eq:eom} is satisfied if $\mathcal{Y}$ satisfies the ``proper-time'' Schr\"{o}dinger equation
\be
i\partial_s \mathcal{Y} = \hat{\mathcal{H}}_h\mathcal{Y},
 \label{eq:27}
\ee
with the boundary condition $\mathcal{Y}|_{s=0} = 1$. We see that $\hat{\mathcal{H} }_h$ generates evolution in the time parameter $s$, hence the name Hamiltonian.

\subsubsection{III.2 (b): WKB Series Expansion}
\label{sec:eikonWKB}

Now, to set up the eikonal expansion, we introduce a WKB representation for the function $\mathcal{Y}$, in the form
\be
\mathcal{Y} \equiv e^{\chi},
  \label{ansatz}
\ee
with $\chi$ expanded as a power series in the coupling
\be
\chi \equiv \sum_{n=1}^\infty \kappa^n \chi_n \label{expansion}.
\ee
In eq.~\eqref{eq:schwingrep} we've already separated off the free propagator factor, so $\chi$ should vanish when $\kappa=0$ and hence the WKB series should indeed start at $\mathcal{O}(\kappa^{1})$. It should then be clear that higher order derivatives of $\mathcal{Y}$ will bring down higher orders in the WKB expansion. Since the background is slowly varying, we then want to retain only the leading order terms in this series. To first order the equation of motion for $\chi_1$ is
\begin{equation} \label{eq:chi1}
\left[i\partial_s + \partial^2 + 2ik^\mu\partial_\mu \right] \chi_1 \;=\; h^{\mu\nu}(x)k_{\mu}k_{\nu},
\end{equation}
and one can continue the hierarchy to find equations of motion for the higher order terms in the WKB series. We then have a systematic expansion suited for studying sub-leading soft-graviton effects.

If we now Fourier transform both sides of eq.~\eqref{eq:chi1} with respect to the center of mass coordinate $X$ we have a simple linear ODE which has solution
\begin{equation}
\kappa\chi_{1}(q,k,s)=-i\kappa h^{\mu\nu}(q)k_{\mu}k_{\nu}\int_{0}^{s}\,ds^{\prime}\,e^{-is^{\prime}(2k\cdot q+q^{2})}.
\end{equation}
and dropping the $\partial^2$ term in the differential operator in \eqref{eq:chi1} - justified by the slowly-varying background assumption - allows us to drop the $q^2$ term in the exponent without altering our results to leading order.

Note that in the language of Feynman diagrams, a graviton-scalar vertex with a transverse-tracless $h_{\mu\nu}$ injecting momentum $q$ is $\kappa h_{\mu\nu}(q)\tau^{\mu\nu}(k,k+q,q)$ where $\tau^{\mu\nu}(k_{1},k_{2},q)=\frac{1}{2}(k_{1}^{\mu}k_{2}^{\nu}+k_{1}^{\nu}k_{2}^{\mu})$. When the incoming matter momentum $k$ is much larger that of the graviton $q$, $k+q\approx k$ we recover the factor $\kappa h^{\mu\nu}(q)k_{\mu}k_{\nu}$ in the above expression. Recalling the preamble to this subsection, we see that this factor is a clear manifestation of eikonal physics---the particle momentum is not changing while interacting with the slowly-varying graviton background.

If we truncate the WKB series at leading order we have that
\begin{equation}
\mathcal{G}_k(X|h) = -i\int_0^{\infty} ds \,  e^{-is(k^{2}+m^{2})+\kappa\sum_{q}e^{iq\cdot X}\chi_{1}(q,k,s)},
\end{equation}
or, written in position space the leading order eikonal approximation to the full Green function is,
\begin{align} \label{eq:PosSpaceG}
\mathcal{G}(x,y|h) &= -i\sum_k e^{ik\cdot(x-y)} \nonumber \\
&\times  \int_0^\infty ds \,  e^{-is(k^2 + m^2) -i\kappa k_{\mu}k_{\nu} \int_0^s ds'\, h^{\mu\nu}(y - 2s'k)} .
\end{align}

To best understand this expression we will use the method of stationary phase to evaluate the integrals. The eikonal phase $\exp(\kappa\chi_{1})$ is a slowly varying function of $s,k$ relative to both the bare propagator factor and the fourier factor. We can then pull it outside the integral and replace its dependence on $s,k$ by those values extremize the oscillatory part of the integrand, \textit{i.e.}
\begin{align}
&s=\sigma/2m \nonumber \\
&k^{\mu} = \frac{m(x-y)^{\mu}}{\sigma},
\end{align}
where the spacetime interval is $\sigma=\sqrt{-(x-y)^{2}}$. Substituting in these values, we obtain the intuitive expression for the full greens function
\begin{equation}
\mathcal{G}(x,y|h)=e^{\frac{i}{2}\kappa\int d^{4}z \tau^{\mu\nu}(z)h_{\mu\nu}(z)}G_{0}(x,y),
\end{equation}
where
\begin{equation}
\tau^{\mu\nu}(z) = -\frac{p^{\mu}p^{\nu}\sigma}{m}\int^{1}_{0}d\tau \,\delta^{4}(z-X_{\textrm{cl}}(\tau)),
\end{equation}
is precisely the stress-energy density for a classical massive relativistic point particle following the world line $X_{\textrm{cl}}(\tau)=y+(x-y)\tau$. As promised, the leading order eikonal approximation for the greens function on a slowly varying background metric perturbation is given by the free Green function multiplied by an eikonal phase describing the phase accumulated along the classical worldline connecting the two spacetime points.

In summary, our main result is then that all effects of soft gravitons on a scalar matter system are described by the composite generating functional
\begin{equation}
\mathscr{Z}_S[{\bf J}] = \mathcal{F}_{S}[\delta_{\mathbf{h}}]\; Z[J|h] \; Z^{*}[J'|h']
\bigg|_{h,h'=0}
\end{equation}
where
\begin{equation}
Z[J|h]=e^{-\frac{i}{2}\int d^{4}xd^{4}y\, J(x)\mathcal{G}(x,y|h)J(y)}
\end{equation}
and the Green function has the simple form
\begin{equation}\label{eq:dressedpropagator}
\mathcal{G}(x,y|h)=e^{\frac{i}{2}\kappa\int d^{4}z \tau^{\mu\nu}(z)h_{\mu\nu}(z)}G_{0}(x,y).
\end{equation}
This object, the eikonal approximated composite generating functional, generates density matrix propagators which provide expressions for the reduced density matrix of the matter system in which the state of the graviton field has been traced out.  There are a variety of potential applications of this general result, which will be the subject of future publications~\cite{Delisle2018}; in the following section we study scattering problems.


\section{IV: Scattering Problems}
\label{sec:Smatrix}


In this section we consider the specific example of scattering between matter fields and soft gravitons. Quite apart from the questions noted in the introduction, the interest here is that scattering problems are formulated over an infinite times, and so infinite wavelength gravitons are present.  The methods developed above are very naturally adapted to such problems, as we now see.

In conventional scattering problems, pure states evolve from the asymptotic past to the asymptotic future via the S-matrix. In what follows we will describe how reduced density matrices evolve over the same spatiotemporal region via a ``composite S-matrix". Consider an initial product state of two systems $Q$ and $X$ written in the `in' basis, $\rho_{Q}(\alpha,\alpha')\rho_{X}(a,a')$.  The `out' density matrix (which is generically not a product state) is related to the `in' density matrix through the S-matrix,
\begin{eqnarray}
\rho_{X,Q}(b,b';\beta,\beta') \;&=&\; \sum_{\alpha a, \alpha' a'}S_{\beta b,\alpha a}S^{*}_{\beta' b',\alpha' a'} \nonumber \\
&& \qquad \times \; \rho_{Q}(\alpha,\alpha') \; \rho_{X}(a,a') \;\;\;\;
\end{eqnarray}
If we then trace over system $X$ (so that we consider $X$ to be an ``environment"), the evolution of the reduced density matrix for $Q$ can be written
\begin{equation}\label{eq:densityMatrixScaterring}
\rho_{Q}(\beta,\beta')=\sum_{\bm{\alpha}}\mathscr{S}_{\bm{\beta},\bm{\alpha}}\,\rho_{Q}(\alpha,\alpha'),
\end{equation}
where the {\it composite S-matrix} is defined as
\begin{equation}\label{eq:compSmatrixdefn}
\mathscr{S}_{\bm{\beta},\bm{\alpha}}=\sum_{b}\sum_{a,a'} S_{\beta b,\alpha a}S^{*}_{\beta' b,\alpha' a'} \; \rho_{X}(a,a')
\end{equation}

To understand the process of decoherence during any scattering process we must compute this object. In the following we will first derive a ``composite scattering functional" $\mathscr{S}$ from the composite generating functional $\mathscr{Z}[\mathbf{J}]$. The computation is analogous to the standard LSZ procedure in quantum field theory for deriving the S-matrix generating functional~\cite{weinbergQFT}. From this we then derive the result we need for the composite S-matrix $\mathscr{S}_{\bm{\beta},\bm{\alpha}}$.  We are then able to derive the leading eikonal result for the `out' state predicted by the composite S-matrix, and compare this to recent results in the literature derived via diagrammatic methods.

\subsection{IV.1: Composite Scattering functional $\mathscr{S}$ }
\label{sec:SmatrixD}

Recall that the Lehmann-Symanzik-Zimmerman (LSZ) procedure \cite{weinbergQFT} for computing the usual scattering operator $S$ from a generating functional is compactly expressed by the formula
\begin{equation}\label{eq:smatrix}
S=\,:e^{\int d^{4}x \; \phi_{\textrm{in}}(x)G_{0}(x)^{-1}\frac{\delta}{\delta J(x)}}:\,Z[J]\bigg|_{J=0}.
\end{equation}
Again, $G^{-1}_0(x)$ is the inverse free Klein-Gordon Green function in (\ref{G-0}). The `in' field $\phi_{\textrm{in}}$ satisfies the free Klein-Gordon equation
\begin{equation}
G^{-1}_0(x) \phi_{\textrm{in}}(x)=0,
\end{equation}
and is related to the full field $\phi$ via the weak asymptotic limits
\begin{equation}
\lim_{x^{0}\rightarrow-\infty}\big[\langle p|\phi(x)|q\rangle-\langle p|\phi_{\textrm{in}}(x)|q\rangle\big]=0,
\end{equation}
in which $| p \rangle, | q \rangle$ are arbitrary states of the system. We assume that the field is renormalized such that the pole in the two-point function is at $-p^{2}=m^{2}$ and the residue is one, so we do not need to carry around factors of the field strength renormalization. The scalar field has the expansion in positive and negative frequency parts
\begin{equation}
\phi_{\textrm{in}}(x)= \phi^{+}_{\textrm{in}}(x)+\phi^{-}_{\textrm{in}}=\int d^{3}p \,\psi_{p}(x)a_{p} + h.c.\,,
 \label{phiIn}
\end{equation}
where the creation/annihilation operators have the commutation relation normalized as $[a_{p},a^{\dagger}_{k}]=\delta^{3}(p-k)$ and states are normalized as $|p\rangle\equiv a^{\dagger}_{p}|0\rangle$, implying the wavefunctions are normalized as
\begin{equation}
\psi_{p}(x)=\frac{e^{-ipx}}{(2\pi)^{3/2}\sqrt{2E_{p}}},
 \label{psip}
\end{equation}
where $E_{p}=\sqrt{|\bm{p}|^{2}+m^{2}}$ is the energy of a particle.

The elements of the S-matrix are given as usual by $S_{\beta,\alpha}={}_{\textrm{in}}\langle\beta|S|\alpha\rangle_{\textrm{in}}$. The expression for the S-matrix we're using is convenient because of the explicit dependence on the generating functional, an object which we can write in path-integral expression form.  It must be noted that, as written, eq.~\eqref{eq:smatrix} can only generate S-matrix elements with $\phi$ particles in the in/out states.  It will soon be clear that for our purposes this won't cause any problems.

If we wanted to compute the product of S-matrix elements $S_{\beta,\alpha}S^{*}_{\beta',\alpha'}$ we would consider two copies of the Hilbert space $\mathcal{H}\otimes\mathcal{H}$ and take matrix elements of $S\otimes S^{\dagger}$, where the operators in the first $S$ commute with those in the second. Dropping the arguments of the functions we write this as
\begin{equation}\label{eq:ssmatrices}
S\otimes S^{\dagger} = \,:e^{\int  \phi_{\textrm{in}}G_0^{-1}\frac{\delta}{\delta J}}:\otimes \,:e^{\int\phi'_{\textrm{in}}G_0^{-1}\frac{\delta}{\delta J'}}:\,Z[J]Z^{*}[J']\bigg|_{\mathbf{J}=0}.
\end{equation}

Now we've already shown in the derivation of the composite generating functional that we can incorporate the effects of soft gravitons in some process by introducing a background metric perturbation $h_{\mu\nu}$, acting with the soft graviton influence functional operator $\mathcal{F}_S(\delta_{h})$ on the forward and reverse Keldysh amplitudes, and then setting $h_{\mu\nu}\rightarrow 0$. The influence functional operator not only `dresses' the process with soft internal gravitons, it also accounts for the eventual trace over soft graviton brehmstrahlung by introducing correlations between the forward and reverse Keldysh paths. Indeed we obtained the composite generating functional by doing precisely this on $Z[J]$, in eqtns. (\ref{ZsFs}) and (\ref{Z-ZJ}).

The form of the composite scattering operator $\mathscr{S}$ then follows immediately; we have:
\begin{align}\label{eq:compositeSmatrixOperator}
\mathscr{S} &= \mathcal{F}[\delta_{\mathbf{h}}]\,S[h]\otimes S^{\dagger}[h']\bigg|_{\mathbf{h}=0} \nonumber \\
&= \,:e^{\int  \phi_{\textrm{in}}G_0^{-1}\frac{\delta}{\delta J}}:\otimes:e^{\int  \phi'_{\textrm{in}}G_0^{-1}\frac{\delta}{\delta J'}}:\mathscr{Z}[\mathbf{J}]\bigg|_{\mathbf{J}=0}
\end{align}
where $S[h]$ is the scalar S-matrix evaluated on a background metric perturbation $h_{\mu\nu}$.

The scattering operator, or composite scattering functional $\mathscr{S}$, is in some ways a rather peculiar object - it describes scattering, but is non-unitary because it also incorporates the loss of information entailed by the averaging over the gravitons modes. To properly understand its properties we need to compute its matrix elements between in and out matter states.

\subsection{IV.2: Composite S-matrix elements}
\label{sec:Smatrixelem}

By taking matrix elements of the functional operator $\mathscr{S}$ in the double copied Hilbert space we obtain the desired matrix elements $\mathscr{S}_{\bm{\beta},\bm{\alpha}}$. We will start from $\mathscr{S}$ as given in  eq.~\eqref{eq:compositeSmatrixOperator} to compute these matrix elements.

First we briefly review the derivation of the standard S-matrix as defined in eq.~\eqref{eq:smatrix}; and then we see how the parallel derivation works when deriving the composite S-matrix.

\subsubsection{IV.2 (a) Bare S-matrix elements}
 \label{sec:bareS}

\begin{figure}
\includegraphics[width=3.2in]{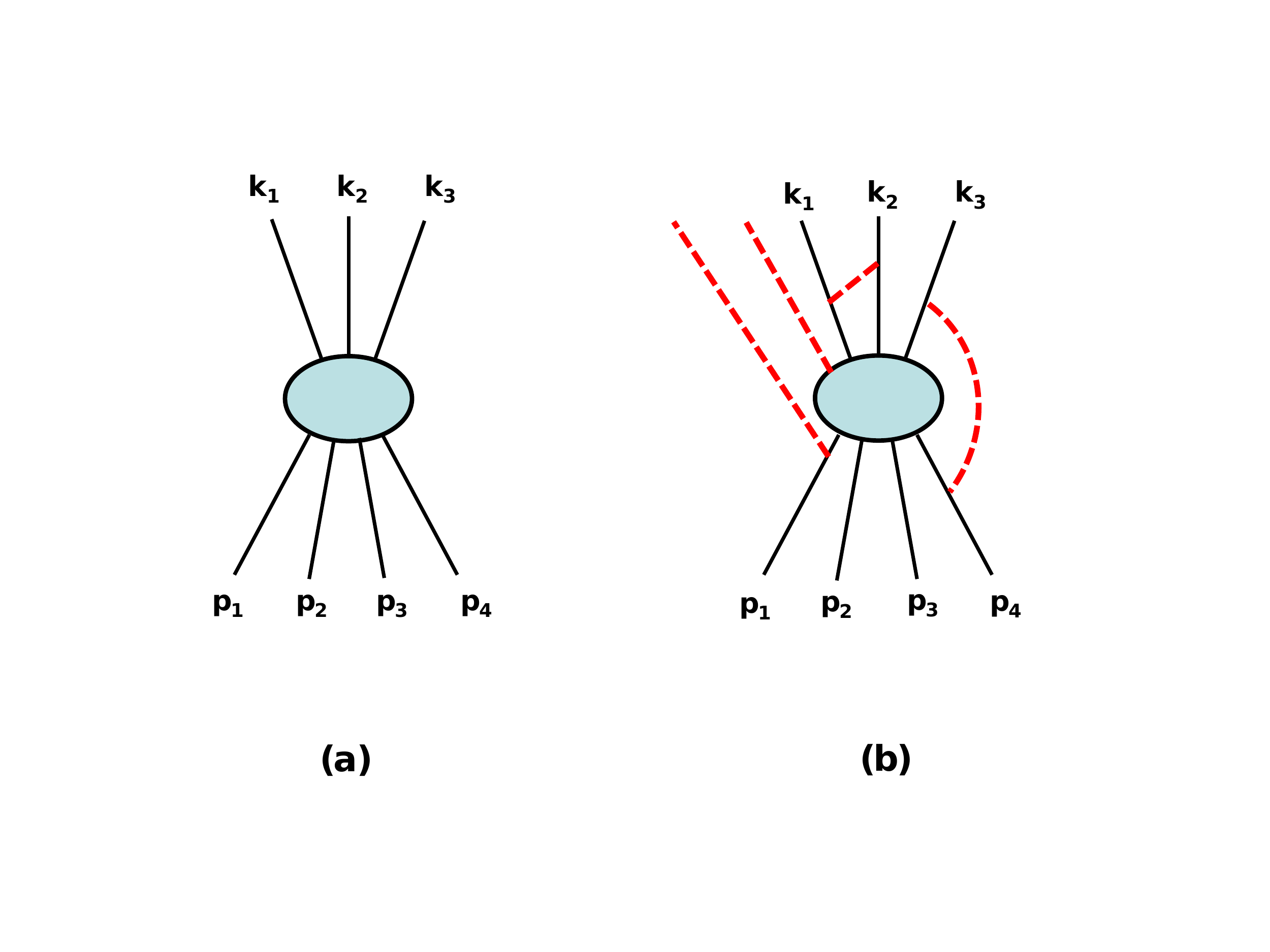}
\caption{\label{Fig:scatt} The scattering processes considered here. In (a) we see the scattering without gravitons, where scalar states $|\alpha\rangle=|p_{1}...p_{n}\rangle$, shown in black, scatter to $|\beta\rangle=|k_{1}...k_{m}\rangle$; the blue oval represents the scattering matrix $S_{\beta\alpha}$. In (b) gravitons are included, in red; the asymptotic graviton states are soft, with $|q| < \Lambda_o$.   }
\end{figure}

In a typical scattering amplitude calculation in conventional field theory (see Fig. \ref{Fig:scatt}) one considers ``$n\rightarrow m$ scattering" between, in our case, scalar states $|\alpha\rangle=|p_{1}...p_{n}\rangle, |\beta\rangle=|k_{1}...k_{m}\rangle$. These states are just those given above, in eqns. (\ref{phiIn}) and (\ref{psip}). There are no other particles or fields considered in the problem - all scattering results from potentials, and is unitary.

One accordingly assumes the existence of scattering states which are approximate eigenstates of the fully interacting Hamiltonian that look like eigenstates of the free Hamiltonian in the asymptotic future and past. The justification is that ultimately one should be working with wavepackets which are sufficiently well separated in the far future and past that they are essentially non-interacting. Scattering states then look like eigenstates of the free Hamiltonian---they are states with definite particle number.


Taking the matrix elements of the scattering operator in eq.~\eqref{eq:smatrix}, we immediately have
\begin{equation}
S_{\beta,\alpha}=\langle 0|a_{k_{1}}\dots a_{k_{m}}e^{\int \phi^{-}G_0^{-1}\frac{\delta}{\delta J}}e^{\int \phi^{+}G_0^{-1}\frac{\delta}{\delta J}}a^{\dagger}_{p_{1}}\dots a^{\dagger}_{p_{n}}|0\rangle.
\end{equation}
Commuting the creation/annihilation operators through the S-matrix we obtain the standard LSZ expression in terms of the amputation and on-shell restriction of the correlation function
\begin{align}
S_{\beta,\alpha}&\equiv S_{\beta,\alpha}[\delta_{J}]Z[J]\bigg|_{J=0} \nonumber \\
&=\int d^{3}y_{1}\dots d^{3}y_{m}\,\psi^{*}_{k_{1}}(y_{1})\dots \psi^{*}_{k_{m}}(y_{m}) \nonumber \\
&\times\int d^{3}x_{1}\dots d^{3}x_{n}\,\psi_{p_{1}}(x_{1})\dots \psi_{p_{n}}(x_{n}) \nonumber \\
&\times G^{-1}_{0}(y_{1})\dots G^{-1}_{0}(y_{m})G^{-1}_{0}(x_{1})\dots G^{-1}_{0}(x_{n}) \nonumber \\
&\times\frac{\delta}{\delta J(y_{1})}\frac{\delta}{\delta J(y_{m})}\dots \frac{\delta}{\delta J(x_{1})}\frac{\delta}{\delta J(x_{n})}Z[J]\bigg|_{J=0}.
\end{align}

From this expression one can compute any S-matrix element between these massive particle states, given an expression for the generating functional (which is typically evaluated as a perturbative series in powers of the coupling constants).

However, let us now note that if one now includes soft gauge excitations like soft gravitons in the scattering calculations, as in/out states along with the massive particles, then our basic assumption of free particle asymptotic states no longer valid. The gravitons have arbitrarily long wavelength, and cannot then be disentangled from the asymptotic matter states. As is well known, this is an essential feature of the infrared divergences in the problem.

\subsubsection{IV.2 (b) Composite S-matrix elements}
 \label{compS}

Now let us consider composite S-matrix elements. To compute such composite S-matrix elements we take matrix elements of the operator  eq.~\eqref{eq:compositeSmatrixOperator} in the basis of the double copy of the Hilbert space $|p_{1}\dots p_{n}\rangle\otimes|p'_{1}\dots p'_{n}\rangle$, to get
\begin{eqnarray}\label{eq:compositeSmatrixelements}
\mathscr{S}_{\bm{\beta},\bm{\alpha}}&=&\mathcal{F}[\delta_{\mathbf{h}}] \bigg( S_{\beta,\alpha}[\delta_{J}]Z[J|h] \nonumber \\
&& \qquad\qquad \times \; S^{*}_{\beta',\alpha'}[\delta_{J'}]Z^{*}[J'|h'] \bigg)\bigg|_{\mathbf{J},\mathbf{h}=0} \qquad
\end{eqnarray}

\begin{figure}
\includegraphics[width=3.2in]{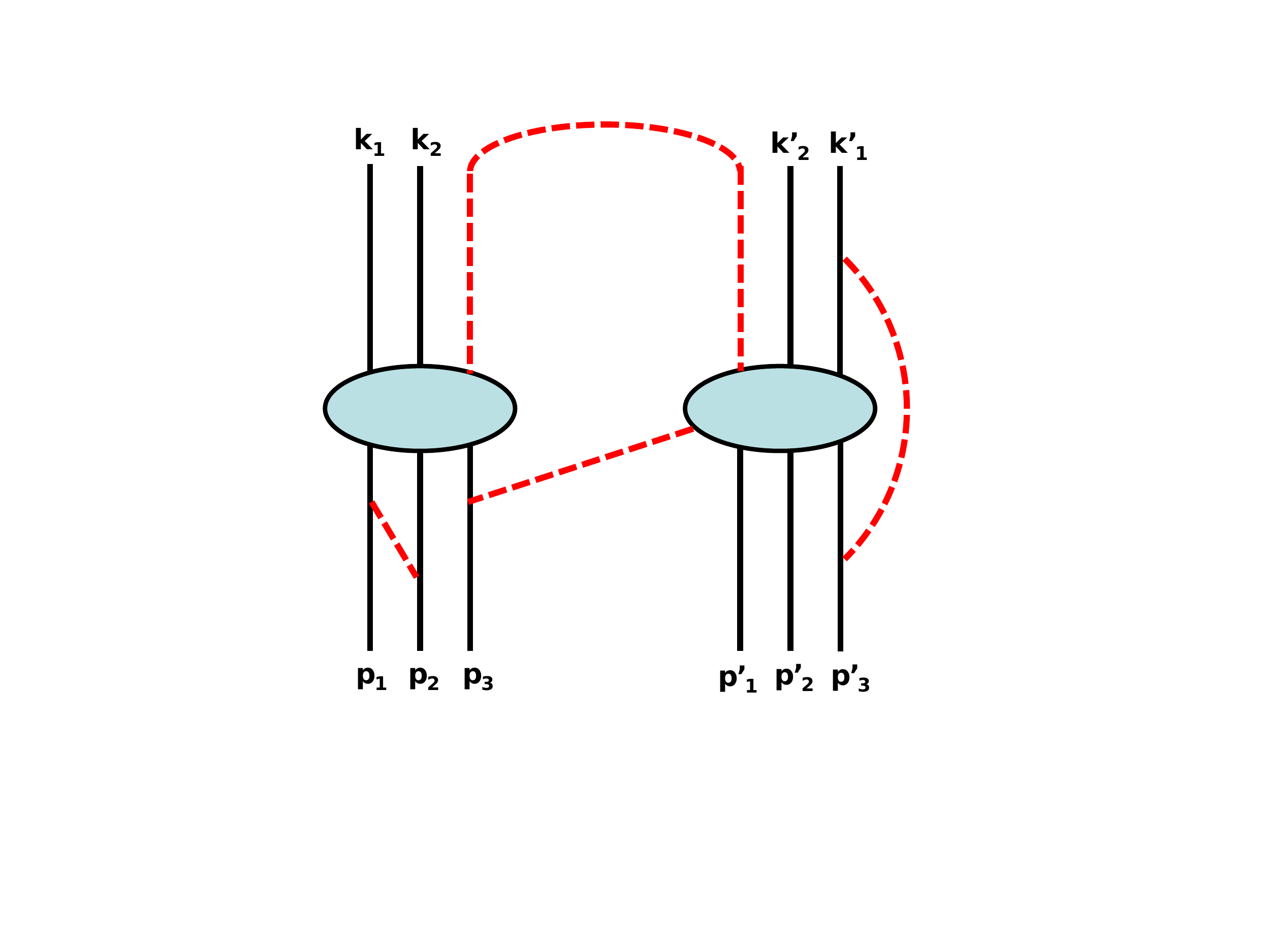}
\caption{\label{Fig:compS} A process contributing to the composite S-matrix $\mathscr{S}_{\bm{\beta},\bm{\alpha}}$, where in the figure $\bm{\alpha} = (p_1, p_2, p_3; p'_1, p'_2, p'_3)$, and $\bm{\beta} = (k_1, k_2; k'_1, k'_2)$. The functional integration over the gravitons (shown in red) includes both graviton exchange between massive particles (shown in black) and gravitons emitted to/absorbed from infinity. }
\end{figure}

Let us take a moment to properly understand this equation, a contribution to which is depicted in Fig. \ref{Fig:compS}. As noted above, we now have arbitrarily long-wavelength gravitons in the problem, which cannot properly be regarded as free particle states. Note, however, that a state describing a definite number of matter particles propagating on a very long-wavelength configuration of the metric perturbation is approximately free - not because the gravitons are well separated from the matter, but because there is very limited momentum exchange with the matter.

In our definition of the composite S-matrix eq.~\eqref{eq:compSmatrixdefn} the environment ``$X$'' is the metric perturbation field $h_{\mu\nu}(x)$. Its states, indexed by $a,b,a',b'$, are not states of definite soft-graviton number; instead we assume a basis of Schrodinger states $\{|h_{ij}\rangle\}$ corresponding to states with definite slowly-varying field configuration $\hat{h}_{ij}(x)|h_{ij}\rangle=h_{ij}(x)|h_{ij}\rangle$. We are using this basis rather than a Fock basis.

We saw in eqtn. (\ref{Z-JJG}) that the generating functional for a non-interacting scalar field living on a slowly varying background metric perturbation can be be written as a simple Gaussian integral, in which the free scalar propagator is replaced by the propagator on a background metric perturbation. We can evaluate the action of $S_{\beta,\alpha}[\delta_{J}]$ on $Z[J|h]$ in (\ref{eq:compositeSmatrixelements}) in precisely the same way, as a generalization of what we would have if there were no background.


With these remarks in mind, let us now consider the term
\begin{equation}
S_{\beta,\alpha}[h] \;=\;  S_{\beta,\alpha}[\delta_{J}]Z[J|h]|_{J=0},
\end{equation}
appearing in eqtn.~\eqref{eq:compositeSmatrixelements}; this describes the S-matrix in the presence of the slowly-varying field $h^{\mu\nu}(x)$.  There will be both internal processes inside the ``blue oval", coming from, eg., a $\phi^{4}$ term in the Lagrangian, or perhaps be mediated by another field; and then there are external matter lines. We see that the effect of $h^{\mu\nu}(x)$ on any diagrams for the matter field will be to ``dress" the scalar propagators according to the eikonal result eq.~\eqref{eq:dressedpropagator}.

In what follows we will make a very simple approximation for the internal scalar propagators - we will assume they can be replaced by the bare propagator $G=\mathcal{G}|_{h=0}$. This apparently drastic simplification is actually equivalent to the assumption that is made in a diagrammatic IR treatment of soft graviton processes\cite{weinberg65}, where soft boson lines are assumed to attach to external legs but not to the internal ``hard process".  All of the dependence on $h_{\mu\nu}$ is then in the external legs of the diagram.

When acting with $G_0^{-1}$ on outgoing external lines (thereby ``amputating" them) we then have contributions from each (outgoing) leg, of form
\begin{equation}
\int d^{4}y \; e^{iky} G^{-1}_{0}(y)\mathcal{G}(y,z|h) \;=\;e^{ikz}e^{i\frac{\kappa}{2}\int d^{4}w \; h^{\mu\nu}(w)\tau_{\mu\nu}(w)},
\end{equation}
where again $\tau_{\mu\nu}(w)=-mU_{\mu}U_{\nu}\int_{0}^{\infty}ds\,\delta^{4}(w-z-sU)$ is the eikonal result for the stress-energy for a scalar excitation whose four-momentum is $p^{\mu}=mU^{\mu}$. Note that $\mathcal{G}(y,z|h)$ comes from the functional derivatives of $Z[J|h]$, with the relevant scattering vertex labeled by $z$.  There is an analogous contribution for ingoing lines, viz.,
\begin{equation}
\int d^{4}x e^{-ikx} G^{-1}_{x}\mathcal{G}(z,x|h) \;=\; e^{-ikz}e^{i\frac{\kappa}{2}\int d^{4}w \; h^{\mu\nu}(w)\tau_{\mu\nu}(w)},
\end{equation}
where for ingoing lines the stress tensor is of the form $\tau_{\mu\nu}(w)=-mU_{\mu}U_{\nu}\int_{0}^{\infty}ds\,\delta^{4}(w-z+sU)$. In the absence of a slowly varying background metric the external leg would be straightforwardly amputated (\textit{i.e.}, the inverse propagator acting on the propagator would yield a delta function), however when the background is treated via this eikonal approximation we obtain an additional eikonal phase for each leg. The analogous expressions for outgoing/ingoing lines on the return Keldysh path are obtained by taking the complex conjugates of these expressions.

Provided the scattering occurs in a region around the origin much smaller than $\Lambda^{-1}$, we can make the approximation that $z\approx0$ within the eikonal phases since the slowly varying field $h_{\mu\nu}(w)$ is indifferent to such a translation. With this approximation the eikonal phases all factor out of the scattering amplitude.

For a standard S-matrix computation we then obtain the eikonal result
\begin{equation}\label{eq:eikonalCompositeSmatrix}
S_{\beta,\alpha}[h]\;\;=\;\; e^{i\frac{\kappa}{2}\int d^{4}w \; h^{\mu\nu}(w)\sum_{a}\tau^{a}_{\mu\nu}(w)} \; S^{\Lambda_{0}}_{\beta,\alpha},
\end{equation}
where $a$ runs over all external legs, and $S^{\Lambda_{0}}_{\beta,\alpha}$ is the S-matrix computed with IR cutoff $\Lambda_{0}$. This is an example of the well-known soft-factorization of scattering amplitudes.

If we now consider the full `composite S-matrix' in eq.~\eqref{eq:compositeSmatrixelements}, then by taking the functional derivative and setting $h=0$ we get the very simple result
\begin{equation}
\mathscr{S}_{\bm{\beta},\bm{\alpha}} \;\;=\;\; \mathcal{F}[\sum_{a}\tau^{a},\sum_{a'}\tau^{\prime a'}] \; S^{\Lambda_{0}}_{\beta,\alpha}S^{\prime\Lambda_{0}}_{\beta',\alpha'}{}^{*},
 \label{compS-F}
\end{equation}
for $\mathscr{S}_{\bm{\beta},\bm{\alpha}}$, in which the entire effect of the soft gravitons has been reduced to the {\it sums}, $\sum_{a}\tau^{a}$ and $\sum_{a'}\tau^{\prime a'}$ over all scalar particles, of the the stress-energies from these particles at their asymptotic end-points - these 2 sums are then the arguments of the influence functional in (\ref{compS-F}).

This is a remarkable simplification, given that the influence functional is usually a functional over all the paths in the particle path-integral. The reason is of course that the only effect of soft gravitons here, as incorporated in the influence functional, is to modify the phases of the `classical' straight-line asymptotic paths of the particles follow straight lines

We have thus reduced the problem of the effect of real and virtual soft gravitons on scattering problems to the computation of the influence functional above.  In the next section we  compute this explicitly and investigate the consequences


\section{V. Influence Functional, BMS Noether charges, \& Gravitational Memory}
\label{sec:BMS}


Our goal in this section is limited - we wish to explore the asymptotic properties of the composite scattering matrix, and see how they are influenced by the decoherence functional. One can do much more than this - by looking at $\mathscr{S}_{\bm{\beta},\bm{\alpha}}$ as a function of time, before reaching the asymptotic regime, one can explore the time dynamics of information loss in the system. However, here we confine ourselves to the asymptotic regime. We find that the decoherence functional yields some previous results for this regime, and a new interpretation of them.

In what follows we use the expression derived in the last section for the composite S-matrix to first find the explicit form of the decoherence functional $\Gamma[T,T']$, and then show how it can be rewritten in terms of the asymptotic BMS charges and gravitational memory for the scattering of soft gravitons. Finally, we discuss the implications of these results for the information loss problem.


\subsection{V.1: Form of Influence Functional}
\label{sec:BMS-infl}


In eqtn. (\ref{eq:QGIF}) we derived an explicit form for the influence functional (\ref{F-Th}), which as we recall can be written as $\mathcal{F}[T,T']=e^{i\Psi_{0}+i\Delta}e^{-\Gamma}$, where $\Psi_{0}, \Delta, \Gamma$  are all real. The ``self-gravity'' and dissipation parts $\Psi_{0}[T,T']$ and $\Delta[T,T']$ merely lead to an overall phase shift for the composite S-matrix, which we will ignore here. The more interesting physics is in the decoherence functional $\Gamma[T,T']$ which suppresses coherence in the outgoing state.

It is useful to rewrite the general form of the decoherence functional given in (\ref{eq:QGIF}) as a momentum space integral, viz.,
\begin{equation}
\Gamma[T,T']=\frac{1}{4M_{P}^{2}}\sum_{\sigma=+,\times}\int^{\Lambda_{0}}\frac{d^{3}q}{(2\pi)^{3}}\frac{1}{|\mathbf{q}|}|\epsilon^{\sigma}_{\mu\nu}\delta T^{\mu\nu}(q)|^{2},
\end{equation}
where $\delta T_{\mu\nu}=T_{\mu\nu}-T_{\mu\nu}^{\prime}$ is the difference between the forward and return stress-tensors, and the on-shell Fourier transform is used
\begin{equation}
T_{\mu\nu}(q)=\int d^{4}z e^{i|\mathbf{q}|z^{0}-i\mathbf{q}\cdot\mathbf{z}}\, T_{\mu\nu}(z).
\end{equation}
Written this way it is clear that the decoherence functional is non-negative, and so the influence functional has modulus $|\mathcal{F}|\in [0,1]$. It either leaves the composite amplitudes unchanged, or it suppresses them.

If we now specialize to the scattering problem discussed in the last section, things simplify drastically. The Fourier transform of the stress-tensor for scattering paths is just
\begin{equation}
\tau_{a}^{\mu\nu}(q)=i\eta_{a}m_{a}\frac{U^{\mu}_{a}U^{\nu}_{a}}{q\cdot U_{a}}
\end{equation}
where $\eta_{a}=\pm1$ depending on whether the index $a$ refers to an outgoing or ingoing particle. Substituting this expression into our decoherence functional we obtain
\begin{widetext}
\begin{equation}
 \label{eq:decohFun}
\Gamma[\sum_{a}\tau^{a},\sum_{a'}\tau^{\prime a'}] \;\;=\;\; \frac{1}{2}\sum_{\sigma=+,\times}\int^{\Lambda_{0}}d^{3}q|\delta B^{\sigma}_{\beta, \alpha}(q)|^{2}
\;\;\;\equiv\;\;\; \frac{1}{2}\sum_{\sigma=+,\times} \int^{\Lambda_{0}}  dq \int d\Omega(\hat{n})\;|\delta B^{\sigma}_{\beta, \alpha}(q, \Omega(\hat{n}) )|^{2}
\end{equation}
\end{widetext}
where $d\Omega(\hat{n})$ is the infinitesimal solid angle in direction $\hat{n}$, and
\begin{equation}\label{eq:softFactor}
B^{\sigma}_{\beta,\alpha}(q)=\frac{1}{(2\pi)^{3/2}\sqrt{2|q|}}M_{P}^{-1}\sum_{a}\eta_{a}\frac{p^{a}_{\mu}p^{a}_{\nu}\epsilon^{\sigma}_{\mu\nu}(q)}{p\cdot q}.
\end{equation}
is the so-called ``soft factor" (see eq. 2.29 in ref.~\cite{weinberg65}). The name comes from the statement that to leading order in the graviton momentum one can add a single soft graviton emission event to an S-matrix element $S_{\beta,\alpha}$ by simply multiplying the original S-matrix element by $B_{\beta,\alpha}$.  This fact is commonly referred to as Weinberg's soft-graviton theorem.

The integral in (\ref{eq:decohFun}) is logarithmically divergent. This divergence means that unless $\delta B^{\sigma}_{\beta, \alpha}(q, \Omega(\hat{n}) ) = 0$ for every angle on the sphere $\hat{n}$ and for each polarization $\sigma$, the decoherence functional diverges and thus the influence functional as well as the composite S-matrix element will vanish.

This result was previously reported in a slightly less general form~\cite{carney17}. These authors used the Weinberg diagrammatic approach to handle the IR divergences, and assumed that the matter in-state was a momentum eigenstate (\textit{i.e.} only $\alpha'=\alpha$ was considered).  If we choose to assume this initial condition as well, we recover their result.

It is actually illuminating to understand the relation between the WKB path integral result here and the derivations of Weinberg \cite{weinberg65}, and other similar recent discussions \cite{stromP,carney17}. Weinberg showed perturbatively that a specific infinite diagrammatic sum - of soft factors from all diagrams in which soft boson lines (both virtual and real) are inserted into a ``hard'' process -  will exponentiate in a manner that renders scattering rates finite.

The WKB expansion - which is a natural and systematic approximation for quantum systems propagating on slowly varying backgrounds - already yields to lowest order a decoherence functional in which the soft factors are exponentiated. The next correction to leading WKB then leads to sub-dominant corrections to the Weinberg result; and so on. The decoherence functional also allows a useful interpretation of the diagrammatic expansions.  For every momentum $q$ with $|q|\ll \Lambda_{0}$, and for each polarization $\sigma$, the decoherence functional \eqref{eq:decohFun} compares the soft factors for the forward and return Keldysh paths. If these factors are identical, ie., if an emitted soft graviton  $|q, \sigma \rangle$ does not carry information discerning between the two processes, then that mode does not contribute to decoherence. Otherwise the soft-factor is different for the two paths, and by eq.~\eqref{eq:decohFun} there is a contribution to the decoherence functional.


\subsection{V.2: BMS Charges and Gravitational Memory}
\label{sec:BMS-infl2}


Let us now turn to several other ways of expressing the decoherence functional.  The first will involve the so-called with the Bondi-Metzner-Sachs (BMS) charges, associated with
the Bondi-Metzner-Sachs group of supertranslation symmetries~\cite{BondiMetzner1962, sachs1962}. The second will involve what is called gravitational memory. The connection between information loss, BMS symmetries, and gravitational memory has been the topic of a number of recent papers \cite{stromN,pasterski15,laddha2015}, sometimes described in terms of an ``infrared triangle". We will see that the connection to the decoherence functional gives further illumination of these relationships.

In what follows we discuss both BMS charges and gravitational memory, in each case by first briefly recalling what these terms refer to, and then showing how the decoherence functional can be understood in terms of them.


\subsubsection{V.2 (a): BMS Charges}
\label{sec:BMS-infl2a}


There is a large literature on Bondi-Metzner-Sachs (BMS) charges and the BMS group (see refs.~\cite{stromN, ashtekarBook} and refs. therein); here we will simply make the connection with our results on decoherence.

The BMS group is the group of diffeomorphisms which act on null infinity to map one asymptotically flat solution to the Einstein equations to another, potentially physically inequivalent one. A subset of the generators of this group are the six Lorentz generators; their action is well understood in quantum field theory and will not be further discussed. The more interesting part of the group are the remaining supertranslation transformations, of which there are infinitely many.

Supertranslations are defined by functions $f(z,\bar{z})$ on the sphere. In retarded Bondi coordinates $(u,r,z,\bar{z})$ the supertranslation vector field on future null-infinity is
\begin{equation}
\zeta=f\partial_{u}-\frac{1}{r}(D^{\bar{z}}f\partial_{\bar{z}}+D^{z}f\partial_{z})+D^{z}D_{z}f\partial_{r},
\end{equation}
where $D_{z}$ is the covariant derivative with respect to the unit sphere metric $\gamma_{z\bar{z}}=2(1+z\bar{z})^{-2}$. They are a generalization from the four standard global translations to a group of angle-dependent translations in the retarded time $u$. There is an analogous expression in advanced Bondi coordinates for the supertranslation vector field on past null infinity. In general one can perform independent transformations on future null infinity $\mathscr{I}^{+}$ and past null infinity  $\mathscr{I}^{-}$ and thus the BMS group can be written as the direct product $BMS^{+}\times BMS^{-}$. It has been demonstrated that the ``diagonal'' subgroup in which the same function $f(z,\bar{z})$ is used to simultaneously supertranslate both $\mathscr{I}^{+}$ and $\mathscr{I}^{-}$ is a symmetry of quantum gravity linearized about Minkowski space, and the associated Noether charges $Q_{f}$ have been constructed. For pure gravity coupled to massless scalar matter the supertranslation charge on $\mathscr{I}^{+}$ is
\begin{equation}\label{eq:supercharge}
Q_{f}=\frac{1}{4\pi G}\int_{\mathscr{I}^{+}}dud^{2}z\gamma_{z\bar{z}}f\left[T_{uu}-\frac{1}{4}(D^{2}_{z}N^{zz}+D^{2}_{\bar{z}}N^{\bar{z}\bar{z}})\right],
\end{equation}
where
\begin{equation}
T_{uu}=\frac{1}{4}N_{zz}N^{zz}+4\pi G\lim\limits_{r\rightarrow\infty}\left[r^{2}T^{M}_{uu}\right].
\end{equation}
The matter stress-tensor is denoted by $T^{M}_{\mu\nu}$, and $N_{zz}$ is the Bondi news tensor describing outgoing gravitational waves. Again, there is an analogous expression in advanced coordinates on $\mathscr{I}^{-}$. It should be noted that since supertranslations are defined on null infinity $\mathscr{I}^{+}\cup\mathscr{I}^{-}$ and massive particles never reach null infinity, some work must be done to obtain the correct expression for the hard supertranslation charge in a theory with asymptotically stable massive particles~\cite{laddha2015}. The supertranslation charge can be understood as the sum of ``hard'' and ``soft'' contributions. The first term in eq.~\eqref{eq:supercharge} measures the weighted energy flux through $\mathscr{I}^{+}$ and is called $Q^{\textrm{hard}}_{f}$ while the second term is linear in the zero frequency graviton creation operator and hence called $Q^{\textrm{soft}}_{f}$.

There is a particular choice of the function $f$ which picks out a single angle $\hat{n}$ on the asymptotic sphere and a single graviton polarization $\sigma$ (see~\cite{laddha2015}); with this choice, scattering states of the scalar field are eigenstates of the hard-matter part of the supertranslation charge
\begin{align}
\hat{Q}^{\textrm{hard}}_{\hat{n},\sigma}|\alpha\rangle\;=\; Q^{\textrm{hard}}_{\hat{n},\sigma}(\alpha)|\alpha\rangle
 \label{Qhard1}
\end{align}
with the eigenvalue $Q^{\textrm{hard}}_{\hat{n},\sigma}(\alpha)$ given by
\begin{equation}
Q^{\textrm{hard}}_{\hat{n},\sigma}(\alpha) \;=\;  \sum_{a\in\alpha}\frac{-m_{a}\epsilon_{\mu\nu}^{\sigma}(\hat{n})U^{\mu}U^{\nu}}{-U^{0}_{a}+\hat{n}\cdot \vec{U}_{a}}
 \label{Qhard2}
\end{equation}

We immediately see that we can connect all this with the decoherence functional - one simply rewrites the decoherence functional in (\ref{eq:decohFun}) in the form
\begin{widetext}
\begin{align}\label{eq:decohFunct}
\Gamma[\sum_{a}\tau^{a},\sum_{a'}\tau^{\prime a'}] \;\;=\;\; \frac{1}{4M_{P}^{2}}\bigg(\int^{\Lambda_{0}}\frac{dq}{q}\bigg)\sum_{\sigma=+,\times}\int d\Omega(\hat{n}) \big|\Delta Q^{\textrm{hard}}_{\hat{n},\sigma} - \Delta Q^{\textrm{hard} \prime}_{\hat{n},\sigma} \big|^{2},
\end{align}
\end{widetext}
where $d\Omega(\hat{n})$ is the solid angle area element, and where $\Delta Q^{\textrm{hard}}_{\hat{n},\sigma}$ denotes the difference between the in- and out-state values of   $Q^{\textrm{hard}}_{\hat{n},\sigma}$, the hard charge eigenvalue defined in (\ref{Qhard2})), ie.,
\begin{equation}
\Delta Q^{\textrm{hard}}_{\hat{n},\sigma} = Q^{\textrm{hard}}_{\hat{n},\sigma}(\beta)-Q^{\textrm{hard}}_{\hat{n},\sigma}(\alpha)
 \label{DeltaQH}
\end{equation}
where, as usual, we use the primed symbols refer to the return Keldysh path while the unprimed symbols refer to the forward Keldysh path. Eqtn. (\ref{eq:decohFunct}) expresses the decoherence functional in terms of the ``BMS supertranslation charges". We note that, as before, because of the logarithmic divergence of the integration over $q$, the decoherence functional here diverges unless the difference in hard supertranslation charges is the same on the forward and return Keldysh paths, ie., unless
\begin{widetext}
\begin{equation}
{\bf \bar{Q}}^{\textrm{hard}}_{\hat{n},\sigma}  \;\;\equiv \;\;    (Q^{\textrm{hard}}_{\hat{n},\sigma}(\beta)-Q^{\textrm{hard}}_{\hat{n},\sigma}(\alpha))-(Q^{\textrm{hard}}_{\hat{n},\sigma}(\beta')- Q^{\textrm{hard}}_{\hat{n},\sigma}(\alpha')) \;\;=\;\; 0
 \label{Qh-cons}
\end{equation}
\end{widetext}
which is a kind of ``sum rule" for the scattering process - decoherence will suppress all scattering for which this identity is not satisfied.


\subsubsection{V.2 (b): Ward identities}
\label{sec:BMS-Ward}


The result (\ref{eq:decohFunct}) begs for a physical explanation. Actually it follows from supertranslation charge conservation alone. Recall that the soft theorems are an expression of the Ward identity describing the conservation of supertranslation charge.  The Ward identity can be written in general as the statement that the charge commutes with the Hamiltonian and thus the S-matrix
\begin{equation}
\langle\beta|[\hat{Q}^{\sigma}_{\hat{n}},S]|\alpha\rangle=0 ,
\end{equation}
where $S$ is the usual S-matrix. Decomposing the supertranslation charge into hard and soft parts, assuming the initial state of the graviton field is the vacuum with zero soft charge, and noting that scalar scattering states are eigenstates of hard supertranslation charge, we can rewrite the Ward identity as the soft graviton theorem
\begin{equation}
 \left[Q^{\textrm{hard}}_{\hat{n},\sigma}(\beta)-Q^{\textrm{hard}}_{\hat{n},\sigma}(\alpha)\right]\langle\beta|S|\alpha\rangle=\langle\beta|\hat{Q}^{\textrm{soft}}_{\hat{n},\sigma}S|\alpha\rangle.
\end{equation}

Following a similar line of argument, we can instead write the following equation
\begin{eqnarray}
\langle\beta|S|\alpha\rangle\langle\alpha'| [S^{\dagger},\hat{Q}_{\hat{n},\sigma}]|\beta'\rangle &=&\langle\beta|[\hat{Q}_{\hat{n},\sigma},S]|\alpha\rangle\langle\alpha'|S^{\dagger}|\beta'\rangle \nonumber\\ &=&0.
\end{eqnarray}
Looking at the first equality and again decomposing the charge into hard and soft parts, assuming an initial state with zero soft charge, and using the hard charge eigenvalues we can then write
\begin{widetext}
\begin{align}
\left[(Q^{\textrm{hard}}_{\hat{n},\sigma}(\beta)-Q^{\textrm{hard}}_{\hat{n},\sigma}(\alpha))-(Q^{\textrm{hard}}_{\hat{n},\sigma}(\beta')- Q^{\textrm{hard}}_{\hat{n},\sigma}(\alpha'))\right]S_{\beta,\alpha}S^{*}_{\beta',\alpha'} \;\;=\;\; \langle\beta|\bigg[S|\alpha\rangle\langle\alpha'|S^{\dagger},\hat{Q}^{\textrm{soft}}_{\hat{n},\sigma}\bigg]|\beta'\rangle
\end{align}

As written, the right hand side is a matrix element between states $\beta,\beta'$ of a commutator between operators on the full Hilbert space. We could instead factor the state into hard and soft parts $|\beta\rangle=|\beta_{S}\rangle|\beta_{H}\rangle$. If we do this and trace over the soft part of the outgoing state we obtain the composite Ward identity
\begin{align}
\left[(Q^{\textrm{hard}}_{\hat{n},\sigma}(\beta)-Q^{\textrm{hard}}_{\hat{n},\sigma}(\alpha))-(Q^{\textrm{hard}}_{\hat{n},\sigma}(\beta')- Q^{\textrm{hard}}_{\hat{n},\sigma}(\alpha'))\right]\sum_{\beta_{S}}S_{\beta,\alpha}S^{*}_{\beta',\alpha'}= \sum_{\beta_{S}}\langle\beta_{S}|\bigg[\langle\beta_{H}|S|\alpha_{H}\rangle|\alpha_{S}\rangle\langle\alpha^{\prime}_{S}|\langle\alpha^{\prime}_{H}|S^{\dagger}|\beta^{\prime}_{H}\rangle,\hat{Q}^{\textrm{soft}}_{\hat{n},\sigma}\bigg]|\beta_{S}\rangle.
\end{align}
Since the RHS is the trace of a commutator, by the cyclic property of the trace the RHS vanishes. We have therefore derived the following identity for the composite S-matrix using the supertranslation Ward identity
\begin{equation}
\left[(Q^{\textrm{hard}}_{\hat{n},\sigma}(\beta)-Q^{\textrm{hard}}_{\hat{n},\sigma}(\alpha))-(Q^{\textrm{hard}}_{\hat{n},\sigma}(\beta')- Q^{\textrm{hard}}_{\hat{n},\sigma}(\alpha'))\right]
\sum_{\beta_{S}}S_{\beta,\alpha}S^{*}_{\beta',\alpha'} \;\;=\;\; 0.
 \label{Qsum}
\end{equation}
\end{widetext}

In other words, for a given pair of processes, either the difference in hard supertranslation charges is the same on the forward and return Keldysh paths or the composite S-matrix element vanishes. This is precisely the result we already derived from the decoherence functional, in the form of the sum rule in (\ref{Qh-cons}), but now demonstrated using a composite Ward identity.


\subsubsection{V.2 (b): Gravitational Memory}
\label{sec:gravMem}


Another way of looking at our result for the decoherence functional in \eqref{eq:decohFunct} is in terms of ``gravitational memory". Classically one can use the linearized Einstein equations to compute the evolution of a metric perturbation $h^{\mu\nu}(x)$ far from a source \cite{MTW73}. Thus we can then calculate the change $\Delta h^{\mu\nu}(x)$ in the metric, comparing at times well before and well after any change in the source, in a far field region at some distance $r_0$ from the source of the gravitational waves, which in our case will be from the scattering event.  Any permanent change in the metric, ie., where $\Delta h^{\mu\nu}(x) \neq 0$, is called gravitational memory \cite{zeldo74,braginsky87}.

The change $\Delta h^{\mu\nu}(x)$ has been worked out for a large variety of different sources; given the classical paths considered in the scattering setup here, then it is  well-known~\cite{braginsky87} that such a process will lead to a static change in the transverse-traceless part of the asymptotic metric given by
\begin{widetext}
\begin{equation}
\Delta h^{TT}_{\mu\nu}(\vec{q}) \;\;=\;\; \frac{1}{r_{0}}\frac{1}{16\pi^{2}M_{P}}\bigg(\sum_{j\in\alpha}\frac{p_{j\mu}p_{j\nu}}{q\cdot p_{j}}-\sum_{j\in\beta}\frac{p_{j\mu}p_{j\nu}}{q\cdot p_{j}} \bigg)^{TT}
\end{equation}
Comparing this with our expression for the the hard supertranslation charges we see that we can rewrite the decoherence functional as
\begin{align}\label{eq:decohFunctWithMemory}
\Gamma[\sum_{a}\tau^{a},\sum_{a'}\tau^{\prime a'}]  \;\;=\;\;  4\pi^{2}r_{0}^{2}\bigg(\int^{\Lambda_{0}}\frac{dq}{q}\bigg)\sum_{\sigma=+,\times}\int d\Omega(\hat{n}) \big|\epsilon^{\sigma}_{\mu\nu} \Delta h_{\mu\nu}(\hat{q})-\epsilon^{\sigma}_{\mu\nu} \Delta h^{\prime}_{\mu\nu}(\hat{q})\big|^{2},
\end{align}
where the difference here is taken between the latest time on future null infinity and the earliest time on future null infinity.
\end{widetext}

Since the asymptotic shift in the metric is in principle observable by considering the shifts in the relative positions of asymptotic detectors, we can understand this expression for the decoherence functional in the following way. Suppose we prepare an array of asymptotic detectors with given relative positions. During a scattering event information about the event is radiated away as soft gravitons, which will induce a static shift in the relative positions of the asymptotic detectors.  Since the information about the scattering event is stored as the shift in their relative positions, the scattered matter is entangled with the detectors. Attempts at demonstrating interference phenomena in the outgoing state of the scattered matter will be undermined by this, and the only states which can interfere will be those obtained by a scattering event which induces the same relative shifts in the asymptotic detectors. This explains the vanishing of most of the elements of the composite S-matrix.

\subsection{V.3: Asymptotic decoherence properties}
\label{sec:implications}

Let us now discuss the implications of these results for the very physical question of what form the final state density matrix must take.

In fact it is clear that the vanishing of the composite S-matrix element, unless the ``sum rule" in (\ref{Qh-cons}) is satisfied, implies that the out-state density matrix must have a very restricted form. To see how this works, let us consider two kinds of in-state for the system. One will be a simple product of momentum eigenstates, whereas the other will be a ``Cat state" in which we superpose two simple product states. We then have the following results:

\vspace{2mm}

(i) {\it Simple Product State}:  Our first state will be the kind of state usually assumed in scattering calculations, in which there are no gravitons and where the initial matter state is a product over momentum eigenstates, ie., we have
\begin{equation}
|\alpha_{1} \rangle \;=\;  \prod_j |p_j^{(1)} \rangle
 \label{phiProd}
\end{equation}
so that the incoming reduced density matrix for the matter has the simple form $\rho(\alpha,\alpha')=\delta_{\alpha,\alpha_{1}}\delta_{\alpha',\alpha_{1}}$, and then by eqtns.~\eqref{eq:densityMatrixScaterring} and (\ref{compS-F}), the outgoing reduced density matrix for the matter is
\begin{eqnarray}
\rho(\beta,\beta') \; &=& \;\mathscr{S}_{\beta\beta',\alpha_1\alpha_1}
\nonumber \\
 \; & \sim & \; S^{\Lambda_{0}}_{\beta,\alpha_1}S^{\prime\Lambda_{0}}_{\beta',\alpha_1}{}^{*} \; \delta_{{\bf \bar{Q}}^{\textrm{hard}}_{\hat{n},\sigma}, 0}
 \label{rhoProd}
\end{eqnarray}
where the Kronecker $\delta$-function term imposes the BMS charge conservation sum rule. Thus the final state density matrix will vanish unless $Q^{\textrm{hard}}_{\hat{n},\sigma}(\beta)=Q^{\textrm{hard}}_{\hat{n},\sigma}(\beta')$ for both polarizations $\sigma$ and all angles $\hat{n}$. As noted in refs.~\cite{carney17, strominger2017} this condition is highly restrictive, and except in some pathological cases it is only satisfied when the two states are identical, $\beta=\beta'$. The trace over soft graviton emission has then rendered the outgoing matter density matrix completely diagonal. Contrasting this with standard unitary scattering in which states like (\ref{phiProd}) can certainly evolve into superpositions of products of momentum eigenstates, ie., where $|\alpha_{1}\rangle\rightarrow\sum_{\beta}S_{\beta,\alpha_{1}}|\beta\rangle$, we see that the emission of soft gravitons has led to complete decoherence in the asymptotic limit where the states have moved off to infinity.

\vspace{2mm}

(ii) {\it Cat State}: Suppose the incoming state, instead of being a simple product of momentum eigenstates, is in a Schrodinger's Cat state, ie., a superposition of states like (\ref{phiProd}). The simplest example would be a state of form
\begin{equation}
|\alpha \rangle = \frac{1}{\sqrt{2}}\;(|\alpha_{1}\rangle+ e^{i\phi}|\alpha_{2}\rangle).
 \label{phiCat}
\end{equation}
where the phase $\phi$ is a marker for the relative phase between the two components of this superposition (each being a simple product state like (\ref{phiProd})).

Then, by eq.~\eqref{eq:densityMatrixScaterring}, the outgoing matter density matrix is
\begin{eqnarray}
 \label{eq:2stateSuperpos}
\rho(\beta,\beta') &=& \mathscr{S}_{\beta\beta',\alpha_{1}\alpha_{1}} \;+\;\mathscr{S}_{\beta\beta',\alpha_{2}\alpha_{2}} \nonumber \\
&& \qquad + \; e^{i\phi}\mathscr{S}_{\beta\beta',\alpha_{1}\alpha_{2}} \;+\; e^{-i\phi}\mathscr{S}_{\beta\beta',\alpha_{2}\alpha_{1}} \qquad
\end{eqnarray}
where, as before, the BMS charge conservation condition is built into these terms using the same $\delta$-function as above. We'll assume that we are in a generic situation, where $\alpha_{1}\neq\alpha_{2}$ implies that $Q^{\textrm{hard}}_{\hat{n},\sigma}(\alpha_{1})\neq Q^{\textrm{hard}}_{\hat{n},\sigma}(\alpha_{2})$. In this case, the BMS charge conservation condition
\begin{equation}
Q^{\textrm{hard}}_{\hat{n},\sigma}(\alpha_{2})-Q^{\textrm{hard}}_{\hat{n},\sigma}(\alpha_{1})=Q^{\textrm{hard}}_{\hat{n},\sigma}(\beta)-Q^{\textrm{hard}}_{\hat{n},\sigma}(\beta'),
\end{equation}
has rather interesting implications for the various terms in eq.~\eqref{eq:2stateSuperpos}.

The first two ``diagonal" terms on the RHS of eq.~\eqref{eq:2stateSuperpos} will vanish unless $\beta=\beta'$. This leads to a rather general statement---diagonal density matrix elements scatter into diagonal density matrix elements.  The last two ``interference" terms on the RHS of eq.~\eqref{eq:2stateSuperpos} will also vanish unless the BMS charge conservation condition is fulfilled. Thus, in the asymptotic limit, all interference terms are destroyed unless ${\bf \bar{Q}}^{\textrm{hard}}_{\hat{n},\sigma} = 0$.

This result is actually quite extraordinary. Physically, for the interference term, the condition that ${\bf \bar{Q}}^{\textrm{hard}}_{\hat{n},\sigma} = 0$ is precisely the condition required for the emitted graviton phases as (as encoded in the soft factors) to be the same for the 2 branches of the superposition. This of course is just the condition that the gravitons are not able to distinguish between the two states. This is what one might expect from standard considerations of measurement theory - an environment cannot cause decoherence between 2 states if it cannot distinguish between them. However it is remarkable that the condition for this to be the case is just the sum rule in eqtn. (\ref{Qh-cons}).


\section{V: Conclusions}
\label{sec:fin}


Let us now briefly summarize (i) the results we have found, and (ii) summarize the physical conclusions that emerge from these results, and how they compare with previous arguments.

\vspace{2mm}

{\bf Summary of Results:}  To give a unified discussion of soft graviton problems, we have chosen to use a non-perturbative formalism which allows us to calculate decoherence and information loss for an arbitrary process involving soft graviton emission. We have used this formalism to calculate the decoherence functional for the matter field, and then evaluated this functional for a scattering problem, in the asymptotic limit appropriate to the matter field $S$-matrix. This has allowed us to derive results for a ``composite $S$-matrix", which encodes all information about decoherence in the scattering process.

The decoherence functional $\Gamma$ encodes all information about information loss in any quantum-mechanical process. The functional $\Gamma$ for asymptotic scattering, appearing in the composite $S$-matrix we have derived, can be written either in terms of the BMS asymptotic charges, or the gravitational memory associated with the scattering (compare eqns. (\ref{eq:decohFunct}) and (\ref{eq:decohFunctWithMemory})).  This makes the connection to known results for the BMS asymptotic charges and the gravitational memory function, and shows how the BMS charge conservation condition operates in a very specific way to either impose (or not impose) decoherence on the final states.

\vspace{2mm}

{\bf (ii) Physical Implications:} As we noted in the introduction, there has been widespread disagreement in the literature over the extent to which information loss occurs in scattering processes, with arguments both for \cite{hawking16,hawking17,stromP,carney17}, and against \cite{gabai16,mirbab16,bousso17} the existence of information loss from soft gravitons (or soft photons in QED). Part of the problem is that even when authors start from similar formal frameworks (typically either a coherent state approach, or a perturbative approach), they still do not necessarily arrive at the same conclusions. As an example, one may compare the discussions of refs. \cite{carney17,carney1710} and \cite{stromP} with those in refs. \cite{gabai16,mirbab16}, which arrive at opposite conclusions about information loss starting from the same coherent state formulation of the problem.

Part of the disagreement between different groups stems from the following consideration. Clearly, BMS supertranslation charge conservation requires that classical brehmsstrahlung
radiation is emitted when matter is scattered. A classical charged particle with momentum $p$, receiving an impulse which scatters it to momentum $p'$, creates a gauge field disturbance of form
\begin{equation}
h^{\mu\nu}(k) \;\; \sim \;\; {1 \over |k|} \left[ {p'^{\mu}p'^{\nu} \over k\cdot p'} - {p^{\mu}p^{\nu} \over k\cdot p} \right],
\end{equation}
a sum of contributions from the incoming and outgoing matter momenta.

Now if we choose to dress the incoming matter with radiation which destructively interferes with the $p^{\mu}p^{\nu}/(k\cdot p)$ part of the radiation field, then because soft radiation simply passes through the scattering region \cite{bousso17}, the outgoing state will only contain the
$p'^{\mu}p'^{\nu}/(k\cdot p')$ part of the radiation field. This outgoing state is
also a dressed state, similar to the ingoing state
but with different momentum.  If one works entirely with dressed states then the outgoing radiation field knows nothing about the
incoming matter state, and we get no decoherence; but that is because the incoming radiation field is specifically tuned to get this result! Those groups who do find finite decoherence \cite{hawking16,hawking17,stromP,carney17} do not make this assumption.

It is hardly surprising that changing the initial state changes what one finds for decoherence - depending on the couplings one has, this is a general feature of quantum mechanics. The question, of which conditions should be specified for the incoming matter and radiation fields for the present problem, is then a physical question about state preparation. Many possible scenarios can be imagined here, and in the last section we only considered two of these - we have no space to go through all the possibilities.

One purpose of setting up the formalism described here, which explicitly calculates a decoherence functional, is that questions about information loss can be answered just by looking at this functional, which depends only on the assumed ingoing and outgoing states, and the way in which the average over the gauge field is performed. In our calculations we did not use dressed incoming states; the outgoing radiation field then ``knows" about both the in- and out-states of the matter. Tracing out the radiation then gives the sum rule (\ref{Qsum}), in which the CHANGE in the hard charge must be the same on the forward and return Keldysh paths. If instead we had assumed dressed incoming states, then our sum rule would rather say that the FINAL hard charge must be the same on the forward and return Keldysh paths - a result consistent with those found in refs. \cite{gabai16,mirbab16,bousso17}.

Finally, we emphasize that all our scattering calculations involve asymptotic states, ie., we have only been looking at what happens after decoherence has been given an infinite amount of time to take effect. This is why we get ``all or nothing" results, ie., we find complete decoherence except for a few special states for which our BMS charge conservation condition ${\bf \bar{Q}}^{\textrm{hard}}_{\hat{n},\sigma} = 0$ is satisfied. If we look at the decoherence away from the asymptotic limit, before full decoherence sets in, one gets much more complex results, to be developed elsewhere.

\section{Acknowledgements}

We thank Abhay Ashtekar, Daniel Carney, Laurent Chaurette, and Dominik Neuenfeld for related discussions. This research was supported by NSERC, and by a PGS-D award granted to J.W.-G. by NSERC.


\begin{thebibliography}{99}
\bibitem{hawking76} S.W.Hawking, Phys. Rev. D {\bf14}, 2460 (1976); and Comm. Math. Phys. {\bf 87}, 395 (1982)

\bibitem{unruhW17} W.G. Unruh, R.M. Wald, Rep. Prog. Phys. {\bf 80}, 092002 (2017)

\bibitem{marolf17} D. Marolf, Rep. Prog. Phys. {\bf 80}, 092001 (2017)

\bibitem{stromSG} A. Strominger, JHEP {\bf 07}, 152 (2014);  T. He, V. Lysov, P. Mitra and A. Strominger, JHEP {\bf 05} (2015) 151

\bibitem{hawking16} S.W. Hawking, M.J.Perry, A. Strominger, Phys. Rev. Lett. {\bf 116}, 231301 (2016)

\bibitem{hawking17} S.W. Hawking, M.J.Perry, A. Strominger, JHEP {\bf 05} (2017) 161

\bibitem{HPY14} S. Hyun, S.-A. Park and S.-H. Yi,
JHEP {\bf 06} (2014) 151

\bibitem{ACS14} T. Adamo, E. Casali and D. Skinner,
Class. Q. Grav. {\bf 31}, 225008 (2014)

\bibitem{strom14} T. He, P. Mitra, A.P. Porfyriadis and A. Strominger,
JHEP {\bf 10} (2014) 112

\bibitem{laddha15}  M. Campiglia and A. Laddha, JHEP {\bf 04} (2015) 076; M. Campiglia and A. Laddha, JHEP {\bf 07} (2015) 115

\bibitem{stromN} A. Strominger, ``Lectures on the infrared structure of gravity and gauge theory'',  arXiv:1703.05448

\bibitem{stromP} D. Kapec, M. Perry, A.-M. Raclariu, A. Strominger, Phys. Rev. D {\bf 96}, 085002 (2017)



\bibitem{yennie61} D.R. Yennie, S.C. Frautschi, H. Suura, Ann.
Phys. {\bf 13}, 379 (1961)

\bibitem{weinberg65} S. Weinberg, Phys. Rev. {\bf 140}, B516 (1965)

\bibitem{weinbergQFT} S. Weinberg, ``{\it The Quantum Theory of Fields: Vol. I}", Ch. 13 (Cambridge Univ. Press, 1995).



\bibitem{gabai16} B. Gabai, A. Sever, JHEP {\bf 12} (2016), 095

\bibitem{mirbab16} M. Mirbabayi, M. Porrati, Phys. Rev. Lett. {\bf 117}, 211301 (2016)

\bibitem{bousso17} R. Bousso, M. Porrati, Class. Q. Grav. {\bf 34}, 204001 (2017); see also  R. Bousso, I. Halpern, J. Koeller, Phys. Rev. D {\bf 94}, 064047 (2016)

\bibitem{carney17} D. Carney et al., Phys. Rev. Lett. {\bf 119}, 180502 (2017).



\bibitem{chung65} V. Chung, Phys. Rev. {\bf 140}, B1110 (1965)

\bibitem{kibble68}  T.W.B. Kibble, Phys. Rev. {\bf 175}, 1624 (1968), and refs. therein

\bibitem{faddeevK70}  P.P. Kulish, L.D. Faddeev, Theor. Math. Phys. {\bf 4}, 745 (1970).



\bibitem{carney1710} D. Carney et al., Phys. Rev. D. {\bf 97}, 025007

\bibitem{ChoiAkhoury} S. Choi, and R. Akhoury, J. High Energ. Phys. {\bf 02}, 171 (2018).

\bibitem{strominger2017} A. Strominger, arXiv:1706.07143




\bibitem{decoF} A. Stern, Y. Aharonov, Y. Imry, Phys. Rev. A {\bf 41}, 3436 (1990)

\bibitem{fradkin} E.S. Fradkin, Nucl. Phys. {\bf 76}, 588 (1966); see also G.A. Milekhin, E.S.Fradkin, J.E.T.P. {\bf 18}, 1323 (1964).


\bibitem{Delisle2018} C. DeLisle, J. Wilson-Gerow, P.C.E. Stamp, in preparation.




\bibitem{SdW-GF} J. Schwinger, Proc. Nat. Acad. Sci. {\bf 46}, 1401 (1961), and J. Math. Phys. {\bf 2}, 407 (1961); and L.V. Keldysh, JETP {\bf 20}, 1018 (1965)

\bibitem{schmid82} A. Schmid, J Low Temp. Phys. {\bf 49}, 609 (1982)

\bibitem{bloch} E. Montroll, J. Ward,, Phys. Fluids {\bf 1}, 55 (1958); T. D. Lee, C. N. Yang, Phys. Rev. {\bf 113}, 1165 (1959); and C. Bloch, C. De Dominicias, Nucl. Phys. {\bf 10}, 181 (1959)

\bibitem{ABL64} Y. Aharonov, P.G. Bergmann, J.L. Lebowitz, Phys. Rev. {\bf 134}, B1410 (1964)

\bibitem{schmid87} A. Schmid, Ann. Phys. (NY) {\bf 173}, 103 (1987)

\bibitem{calzetta} There have been many formulations of non-equilibrium dynamics which can be applied in quantum gravity theory, and in which decoherence appears; important examples include A.O. Barvinsky, G.A. Vilkovisky, Phys. Rep. {\bf 119}, 1 (1985), and refs. therein; R.D. Jordan, Phys. Rev. D {\bf 33}, 444 (1986), and Phys. Rev. D {\bf 36}, 3604 (1987); and E. Calzetta, B.L. Hu, Phys. Rev. D {\bf35}, 495 (1987), and Phys. Rev. D {\bf 37}, 2878 (1988). There is considerable overlap between some parts of the work presented herein, and these papers, which we do not have space to review here. The principle difference between the current work and the earlier works is the development here of composite functionals, and their use in the scattering problem.


\bibitem{york72} J.W. York, Phys. Rev. Lett. {\bf 28}, 1082 (1972); G.W. Gibbons, S.W. Hawking, Phys. Rev. D {\bf 15}, 2752 (1977)

\bibitem{wilson17} J. Wilson-Gerow, MSc thesis (UBC, Sept. 2017), and to be published.



\bibitem{feynV63} R.P. Feynman, F.L. Vernon, Ann. Phys. {\bf 24}, 118 (1963)

\bibitem{oniga16} T. Oniga and C.H.-T. Wang. Phys. Rev. D, 93:044027 (2016).



\bibitem{DiracConstraints} P.A.M. Dirac. Can. J. Math, {\bf 2}, 129 (1950); and P.A.M. Dirac. ``{\it Lectures on Quantum Mechanics}", Yeshiva University, NY (1964).


\bibitem{footnote} The addition of interaction terms to the matter action will generally modify the matter stress tensor, but if the interactions are weak (in the sense that the coupling constant is small and perturbation theory is valid) then the additions to the stress tensor will be small compared to the leading order terms discussed here.


\bibitem{BondiMetzner1962} H. Bondi, M. van der Burg, and A. Metzner, Proc. Roy. Soc. {\bf A269} 21 (1962)

\bibitem{sachs1962} R. Sachs, Proc. Roy. Soc.  {\bf A270} 103 (1962)

\bibitem{ashtekarBook} A. Ashtekar, ``{\it Asymptotic Quantization: Based on 1984 Naples Lectures}", (Bibliopolis, edizioni di filosofia e scienze, 1987).




\bibitem{pasterski15} S. Pasterski, ``Asymptotic symmetries and electromagnetic memory", arXiv:1505.00716

\bibitem{stromMemory} A. Strominger, A. Zhiboedov, JHEP {\bf 01}, 086 (2016); arXiv:1411.5745


\bibitem{laddha2015} M. Campiglia, A. Laddha , JHEP {\bf 12} (2015), 094

\bibitem{MTW73} C. Misner, K.S. Thorne, J.A. Wheeler, ``{\it Gravitation}" (Addison-Wesley, 1973)

\bibitem{zeldo74} Ya. B. Zeldovich, A.G Polnarev, Sov. Astro. {\bf 18}, 17 (1974)

\bibitem{braginsky87} V.B. Braginsky, K.S. Thorne, Nature 327.6118 (1987): 123-125

\end{thebibliography}
\end{document}